\documentclass[aps,prb,twocolumn,footinbib,superscriptaddress,longbibliography]{revtex4-2}
\usepackage{rotating}
\usepackage{graphicx}
\usepackage{color}
\usepackage{bm}
\usepackage{amsmath,amssymb}
\usepackage[shortcuts]{extdash}
\usepackage{bbold}
\usepackage[cal=boondoxo]{mathalfa}
\allowdisplaybreaks
\usepackage[bookmarksnumbered,bookmarksopen]{hyperref}
\hypersetup{
   colorlinks=true,       % false: boxed links; true: colored links
   citecolor=blue,        % color of links to bibliography
}

\usepackage{url}
\usepackage{array}
\newcolumntype {L}{>{$}l<{$}}             % switch text <-> math
\newcolumntype {C}{>{$}c<{$}}             % switch text <-> math
\newcolumntype {R}{>{$}r<{$}}             % switch text <-> math
\newcolumntype {s}[1]{@{\hspace*{#1}}}    % space
\newcolumntype {S}[1]{@{\extracolsep{#1}}} % Space
\newcolumntype {e}{@{\extracolsep {0pt}}}
\newcolumntype {E}{@{\extracolsep \fill}}
\newcolumntype {v}{@{\hspace*{2em}}} % vertical line

\newcommand{\marr}[2][@{}l@{}]{\mbox{\renewcommand{\arraystretch}{0.8}%
  \tabcolsep 0pt\begin{tabular}{#1} #2 \end{tabular}}}

\usepackage{braket}
\let\avg\braket

\newcommand* {\vek}[1]{{\bm{\mathrm{#1}}}}
\newcommand* {\vekc}[1]{{\bm{\mathcal{#1}}}}
\newcommand* {\kk}{\vek{k}}
\newcommand* {\frack}[2]{{\Ts\frac{#1}{#2}}}

\newcommand* {\Ts}{\textstyle}
\renewcommand* {\cp}{\mbox{cp}}

\newcommand* {\gtis}{C_\theta} % time inversion symmetry
\newcommand* {\gsis}{C_i}      % space inversion symmetry
\newcommand* {\gstis}{C_{i\theta}} % SIS and TIS combined
\newcommand* {\gsistis}{C_{i \times \theta}} % SIS and TIS
\newcommand* {\Rit}{R_{i \times \theta}} % Rotation group + i + theta
\newcommand* {\inv}[2]{{(\mathrm{#1},#2)}}

\newcommand* {\bohrmag}{\mu_\mathrm{B}}
\let\oplus+

\let\myRe\Re
\let\myIm\Im
\renewcommand{\Re}{\myRe\mathrm{e}\,}
\renewcommand{\Im}{\myIm\mathrm{m}\,}

\graphicspath{{./FIG/}}

\begin{document}

\title{Theory of electric, magnetic, and toroidal polarizations in crystalline solids\\ with applications to hexagonal lonsdaleite and cubic diamond}

\thanks{Dedicated to E.~I. Rashba on the occasion of his 95th birthday.}

\author{R. Winkler}
\affiliation{Department of Physics, Northern Illinois University, DeKalb, Illinois, 60115, USA}
\affiliation{Materials Science Division, Argonne National Laboratory,
Lemont, Illinois 60439, USA}

\author{U. Z\"ulicke}
\affiliation{MacDiarmid Institute, School of Chemical and Physical Sciences,
Victoria University of Wellington, PO Box 600, Wellington 6140, New
Zealand}

\date{\today}

\begin{abstract}
Multipolar order in bulk crystalline solids is characterized by
multipole densities --- denoted as \emph{polarizations} in
this work --- that cannot be cleanly defined using the concepts of
classical electromagnetism.  Here we use group theory to overcome this
difficulty and present a systematic study of electric, magnetic, and toroidal
multipolar order in crystalline solids.  Based on our symmetry analysis,
we identify five categories of polarized matter, each of which is
characterized by distinct features in the electronic band structure.  For
example, Rashba spin splitting in electropolar bulk materials like wurtzite
represents the electric dipolarization in these materials.  We also develop
a general formalism of indicators for individual multipole densities that
provide a physical interpretation and quantification of multipolar
order.  Our work clarifies the relation between patterns of
localized multipoles and macroscopic multipole densities they give
rise to.  To illustrate the general theory, we discuss its
application to polarized variants of hexagonal lonsdaleite and cubic
diamond structures.  Our work provides a general framework for
classifying and expanding current understanding of multipolar order
in complex materials.
\end{abstract}

\maketitle

\section{Introduction}

It is well-known that a proper definition of electric dipolarization
as a bulk property is nontrivial \cite{res94, res10}.  The naive
electromagnetic definition of electric dipolarization as the dipole
moment of a unit cell is unsatisfactory as this quantity generally
depends on the arbitrary choice of a unit cell \cite{mar74}.  Thus
a proper description of the electromagnetic properties of solids
requires tools beyond those supplied by classical electrodynamics.

Important progress has been made by introducing the modern theories
of electric dipolarization and magnetization where geometric phases are
used to quantify these dipole densities (multipole order $\ell = 1$)
independently of the choice of unit cell \cite{kin93, res94, res10, res20}.
Within the modern theory, the electric dipolarization has a clear physical
interpretation relative to a reference state.  However, for systems showing a
spontaneous electric dipolarization, the interpretation and observability of
this quantity have remained ambiguous.  Also, it is a significant challenge to
extend the modern-theory approaches to multipole densities of higher
multipole order $\ell > 1$~\cite{gli22}.

Even before the advent of the modern theories, some early studies did not
make any reference to electromagnetism in their investigation of
dipolarizations in materials, as they recognized how crystal symmetry
allows one to identify crystal structures that permit a bulk electric
dipolarization (so-called polar crystals include pyroelectric and
ferroelectric media \cite{voi10, nye85, lit86}) or a bulk magnetization
(ferromagnetic crystals \cite{tav56, tav58, cor58a}).  According to
Neumann's principle (see Refs.~\cite{voi10, nye85} for seminal
discussions), the crystal classes can be rigorously divided into those that
permit a macroscopic electric dipolarization or magnetization, and those
for which these phenomena are forbidden.  Magnetic crystal classes that
do not permit a magnetization have been generically associated with
antiferromagnetism \cite{cor58a}.

The approach pursued in the present work overcomes the unsatisfactory
electromagnetic definition of electric and magnetic multipole densities that
is inadequate for crystalline solids; we rely entirely on symmetry to extend
the notion of bulk dipolarization and magnetization to electric and
magnetic multipole densities of higher orders $\ell > 1$.  To this end, we
treat the black-white symmetries space inversion symmetry (SIS) and time
inversion symmetry (TIS) on the same footing \cite{lud96}.  Moreover, we
treat electric and magnetic order on the same footing.  Our systematic
theory provides a broader framework for recent efforts to
study electric and magnetic multipolar order in solids \cite{wat18, hay18c,
yat21, bho22, bho22a} and lends itself for wider application in the context
of complex materials~\cite{san09, kur09, fu15, nor15, hay19a, vol21}.  Throughout this work, we focus on systems that are in thermal equilibrium,
thus leaving aside the interesting topic of current-induced multipolar order~\cite{tah22}.

\begin{table}[b]
  \caption{\label{tab:tensor-signat} Signature $s s'$ of multipoles
  of order $\ell$. The four different types of multipoles arising
  for any given $\ell$ are associated with the respective
  irreducible representations $D_\ell^{ss'}$ of the full rotation
  group $\Rit \equiv R \times \gsistis$, where $R \equiv SO(3)$ is
  the proper rotation group.}
  \renewcommand{\arraystretch}{1.2}
  \begin{tabular*}{\linewidth}{cE*{4}{C}}
    \hline \hline \rule{0pt}{3.5ex}
    & \mbox{electric} & \mbox{magnetic}
    & \mbox{\marr{electro- \\ toroidal}}
    & \mbox{\marr{magneto- \\ toroidal}} \\ \hline
    $\ell$ even & ++ & -- & -+ & +- \\
    $\ell$ odd  & -+ & +- & ++ & -- \\ \hline \hline
  \end{tabular*}
\end{table}

\begin{table*}
  \caption{\label{tab:mpole-even-odd} Families of electric,
  magnetic, and toroidal multipole densities
  (\emph{polarizations}) of order $\ell$ permitted by SIS, TIS and
  CIS.  Symmetry operations present (absent) in a given inversion
  group are labeled ``$\bullet$'' (``$\circ$'').  Polarizations that
  are allowed (forbidden) under an inversion group are likewise
  labeled``$\bullet$'' (``$\circ$'').  We also list the signature
  $ss'$ for each family of polarizations, with $s = \pm$ ($s' =
  \pm$) indicating transformation behavior under space inversion $i$
  (time inversion $\theta$).  Each inversion group defines a
  category of polarized matter, as indicated in the last column.}
  \renewcommand{\arraystretch}{1.2}
  \let\mc\multicolumn
  \tabcolsep 0pt
  \let\blt\bullet \let\tms\circ
  \begin{tabular*}{\linewidth}{LECCCs{1.5em}*{8}{C}r}
    \hline \hline \rule{0pt}{2.8ex}
    & & & & \multicolumn{2}{c}{electric} & \multicolumn{2}{c}{magnetic} 
    & \multicolumn{2}{c}{electrotoroidal} 
    & \multicolumn{2}{c}{magnetotoroidal} \\
    & \multicolumn{3}{l}{symmetry} 
    & \ell~\mathrm{even} & \ell~\mathrm{odd}
    & \ell~\mathrm{even} & \ell~\mathrm{odd}
    & \ell~\mathrm{even} & \ell~\mathrm{odd}
    & \ell~\mathrm{even} & \ell~\mathrm{odd} & \\
    \raisebox{0.3ex}[0pt]{\renewcommand{\arraystretch}{1.0}%
    \begin{tabular}[b]{@{}l@{}}
       inversion \\ group
    \end{tabular}}
    & i & \theta & i \theta & ++ & -+ & -- & +- & -+ & ++ & +- & -- &
    \raisebox{0.3ex}[0pt]{\renewcommand{\arraystretch}{1.0}
    \begin{tabular}[b]{@{}r@{}}
       category of \\ polarized matter
    \end{tabular}} \\ \hline
    \gsistis = \{e, i, \theta, i\theta \}
    & \blt & \blt & \blt & \blt & \tms & \tms & \tms & \tms & \blt & \tms & \tms
    & parapolar (PP) \\
    \gtis = \{e, \theta \}
    & \tms & \blt & \tms & \blt & \blt & \tms & \tms & \blt & \blt & \tms & \tms
    & electropolar (EP) \\
    \gsis = \{e, i \}
    & \blt & \tms & \tms & \blt & \tms & \tms & \blt & \tms & \blt & \blt & \tms
    & magnetopolar (MP) \\
    \gstis = \{e, i \theta \}
    & \tms & \tms & \blt & \blt & \tms & \blt & \tms & \tms & \blt & \tms & \blt
    & antimagnetopolar (AMP) \\
    C_1 = \{ e \}
    & \tms & \tms & \tms & \blt & \blt & \blt & \blt & \blt & \blt & \blt & \blt
    & multipolar (MuP) \\ \hline \hline
  \end{tabular*}
\end{table*}

In the following, the term \emph{polarization} refers to a general
realization of bulk multipolar order with $\ell \ge 0$.  Four types
of polarizations --- electric, magnetic, electrotoroidal, and
magnetotoroidal --- are presented in Table~\ref{tab:tensor-signat}.
The \emph{signature} $ss'$ indicates how a polarization behaves
under space inversion (even/odd if $s = +/-$) and time inversion
(even/odd if $s' = +/-$).  The electric (magnetic) polarization of
order $\ell=1$ corresponds to the electric dipolarization
(magnetization), having signature $-+$ ($+-$).  This general
group-theoretical definition of electric and magnetic multipolar
order is independent of the arbitrary choice of a unit cell.  It is
also independent of a material's electrodynamic properties and,
therefore, applies to both insulators and metals.

A comprehensive classification of ways to combine polarizations is
based on their transformation behavior under SIS, TIS, and the
combined inversion symmetry (CIS) represented by
the operations $i$, $\theta$, and $i\theta$, respectively.  Five
distinct inversion groups can be formed from these symmetry
operations, as defined in Table~\ref{tab:mpole-even-odd}, where we
also indicate which polarizations are permitted under these groups.
The only types of polarizations permitted under the full inversion
group $\gsistis \equiv C_i \times C_\theta$ are even\=/$\ell$ electric
polarizations called \emph{parapolarizations}, and we label the
associated matter category \emph{parapolar}.  On the opposite extreme,
the trivial inversion group $C_1$ containing only the identity $e$ as a
symmetry element allows all polarization types, and we label the category
of polarized matter associated with $C_1$ \emph{multipolar}.  Each of
the remaining inversion groups contains strictly one inversion
operation $i$, $\theta$, or $i\theta$ as a symmetry element.  As a
result, only a single type of electric or magnetic polarization is
symmetry-allowed: odd\=/$\ell$ \emph{electropolarizations} for the
time inversion group $\gtis$, odd\=/$\ell$
\emph{magnetopolarizations} for the space inversion group $\gsis$,
and even\=/$\ell$ \emph{antimagnetopolarizations} for the combined
inversion group $\gstis$.  Our unified treatment
reveals a far-reaching correspondence between electric and magnetic
order in crystalline solids.

Our theory enables us to identify measurable \emph{indicators} that
signal the presence of electric and magnetic order in the electronic band
structure.  Some of these indicators are quite familiar, though
their relation to electric and magnetic order was not established
previously.

A classical example for a bulk crystal with a spontaneous electric
dipolarization is wurtzite \cite{pos90, ber97a}.  Wurtzite is also
the classical example for a bulk crystal showing the Rashba effect
\cite{ras59a}.  Generally, the Rashba effect is characterized by a
term $\vek{\alpha} \cdot (\vek{k} \times \vek{\sigma})$, where
$\hbar\vek{k}$ is crystal momentum and the vector $\vek{\sigma}$ of
Pauli spin matrices represents the spin degree of freedom of the
Bloch electrons \cite{ras06}.  By definition, the Rashba effect is
proportional to a polar vector $\vek{\alpha}$.  In confined
geometries, the vector $\vek{\alpha}$ is commonly associated with a
built-in, or external, electric field that controls the magnitude of
the Rashba effect \cite{las85, win03}.  In bulk materials like
wurtzite, no such intuitive picture exists for the vector $\vek{\alpha}$,
and its physical meaning has remained unclear \cite{ras59a, voo96}. 
The Rashba effect exists in all bulk crystal structures that belong to
one of the ten polar crystal classes \cite{nye85}.  We argue that, in
these structures, the vector $\vek{\alpha}$ represents the bulk
dipolarization ($\ell = 1$).  The Rashba effect is thus a measure of the
spontaneous electric dipolarization in bulk wurtzite structures and other
polar crystals.  Similarly, the Dresselhaus term \cite{dre55} in bulk
zincblende structures represents an electric octupolarization ($\ell = 3$),
and Dresselhaus spin splitting provides a measure of the spontaneous
electric octupolarization in bulk zincblende structures.

\begin{table*}[t]
  \caption{\label{tab:symmetries} Magnetic point groups of
electrically and magnetically polarized variations of lonsdaleite
and diamond.  Starting from a pristine crystal structure
(lonsdaleite or diamond) that is compatible with an even-parity
electric polarization ($\ell=2$ in lonsdaleite and $\ell=4$ in
diamond), its symmetries are broken by electric and/or magnetic
polarizations as indicated.  We use an extended Sch\"onflies
notation \cite{lud96} where minor groups with respect to space
inversion $i$, time inversion $\theta$, and their combination
$i\theta$ are denoted by $G [\tilde{G}]$, $G (\tilde{G})$ and $G
\{\tilde{G}\}$, respectively.  The symbols $\gsis$, $\gtis$ and $\gstis$
denote the order\=/2 groups associated with $i$, $\theta$ and
$i\theta$, and $\gsistis \equiv \gsis \times \gtis$ is the full
inversion group.  Expressions on the right-hand side of an "$=$"
sign reveal how the black-white symmetries $i$, $\theta$ and
$i\theta$ are combined with proper rotations \cite{lud96}.}
\renewcommand{\arraystretch}{1.2} \let\hr\hyperref
  \begin{tabular*}{\linewidth}{Ls{0.6em}lCs{0.5em}c*{2}{ELS{0.3em}Ls{0.3em}l}}
\hline \hline \multicolumn{2}{l}{polarization} & ss' &
\makebox[2em][l]{category} & \multicolumn{3}{c}{Lonsdaleite family}
& \multicolumn{3}{c}{Diamond family} \\ \hline
  \ell=4 & electric & ++ & PP & & & & \hr[sec:diamond:e4]{O_h \times
\gtis} & \hr[sec:diamond:e4]{= O \times \gsistis} &
(diamond) \\ & magnetic & -- & AMP & \hr[sec:lons:m4]{D_{6h}
(D_{3h})} & \hr[sec:lons:m4]{= D_6 (D_3) \times \gstis} & &
\hr[sec:diamond:m4]{O_h (O)} & \hr[sec:diamond:m4]{= O \times
\gstis} & \\ \hline
  \ell = 3 & electric & -+ & EP & \hr[sec:lons:e3]{D_{3h} \times
\gtis} & \hr[sec:lons:e3]{= D_6 [D_3] \times \gtis} & &
\hr[sec:diamond:e3]{T_d \times \gtis} & \hr[sec:diamond:e3]{= O [T]
\times \gtis} & (zincblende) \\ & magnetic & +- & MP &
\hr[sec:lons:m3]{D_{6h} (D_{3d})} & \hr[sec:lons:m3]{= D_6 (D_3)
\times \gsis} & & \hr[sec:diamond:m3]{O_h (T_h)} &
\hr[sec:diamond:m3]{= O (T) \times \gsis} & \\ \hline
  \ell = 2 & electric & ++ & PP & \hr[sec:lons:e2]{D_{6h} \times
\gtis} & \hr[sec:lons:e2]{= D_6 \times \gsistis} &
(lonsdaleite) & \hr[sec:diamond:e2]{D_{4h} \times \gtis} &
\hr[sec:diamond:e2]{= D_4 \times \gsistis} & (strain) \\ &
magnetic & -- & AMP & & & & \hr[sec:diamond:m2]{D_{4h} (D_{2d})} &
\hr[sec:diamond:m2]{= D_4 (D_2) \times \gstis} & \\ & electric
$\parallel$ magnetic & -- & AMP & \hr[sec:lons:m2]{D_{6h} (D_6)} &
\hr[sec:lons:m2]{= D_6 \times \gstis} & &
\hr[sec:diamond:m2:qw]{D_{4h} (D_{2d})} & \hr[sec:diamond:m2:qw]{=
D_4 (D_2) \times \gstis} & \\ & electric $\perp$ magnetic & -- & AMP
& \hr[sec:lons:m2]{D_{2h} (C_{2v})} & \hr[sec:lons:m2]{= D_2 (C_2)
\times \gstis} & & \hr[sec:diamond:m2:qw]{D_{2h} (C_{2v})} &
\hr[sec:diamond:m2:qw]{= D_2 (C_2) \times \gstis} & \\ \hline
  \ell = 1 & electric & -+ & EP & \hr[sec:lons:e1]{C_{6v} \times
\gtis} & \hr[sec:lons:e1]{= D_6 [C_6] \times \gtis} & (wurtzite) &
\hr[sec:diamond:e1]{C_{4v} \times \gtis} & \hr[sec:diamond:e1]{= D_4
[C_4] \times \gtis} & \\ & magnetic & +- & MP &
\hr[sec:lons:m1]{D_{6h} (C_{6h})} & \hr[sec:lons:m1]{= D_6 (C_6)
\times \gsis} & & \hr[sec:diamond:m1]{D_{4h} (C_{4h})} &
\hr[sec:diamond:m1]{= D_4 (C_4) \times \gsis} & \\ \hline \hline
\end{tabular*}
\end{table*}

In finite systems such as molecules, only the lowest\=/$\ell$ nonvanishing
electric and magnetic multipoles are well-defined because
higher-order multipoles depend on the choice of origin of the
coordinate systems \cite{jac99, raa05}.  This problem is closely
related to the problem described above, where, for infinite
crystalline solids, even the multipole density of lowest
nonvanishing order cannot be identified with the multipole moment of
a unit cell because this moment depends on the arbitrary definition
of the unit cell.  We show that multipole densities in crystalline
solids must, indeed, be divided into four \emph{families}
representing even\=/$\ell$ and odd\=/$\ell$ electric and magnetic
multipoles.  Within each family, only the lowest-order multipole
density is well-defined.  But even\=/$\ell$ electric (magnetic)
multipole densities can be defined independent of odd\=/$\ell$
electric (magnetic) multipole densities.

Atomic multipoles can act as microscopic building blocks for
macroscopic multipolar order, including higher-order atomic magnetic
multipoles beyond magnetic dipoles \cite{kur08}.  We demonstrate
that the order and orientation of these local multipoles are fixed
by site symmetries that are tabulated for all crystallographic space
groups \cite{hah05, lit13}.  But the order of the local multipoles
proves to not be simply related to the order $\ell$ of the macroscopic
multipole densities they generate.  For example, in diamond structures,
atomic $sp^3$ hybrid orbitals form local electric octupoles whose
configuration results in a macroscopic hexadecapolarization with
$\ell = 4$.

Toroidal order has been viewed as an essential complement to
electric and magnetic multipolar order \cite{art85, dub90, spa08,
pap16, nan16}.  However, the physical significance of toroidal
moments is being debated \cite{fer17}.  Our work clarifies the role
of toroidal order in solids.  Under the full rotation group $\Rit
\equiv R \times \gsistis$, where $R \equiv SO(3)$ is the proper
rotation group, toroidal moments are fundamentally distinct from
electric and magnetic multipoles, where these moments transform
according to different irreducible representations (IRs) of $\Rit$
(Table~\ref{tab:tensor-signat}).  This distinction is lost in a
crystalline environment, where the IRs of $\Rit$ representing
toroidal moments are mapped onto the same finite set of IRs of the
crystallographic point groups as the IRs of $\Rit$ representing
electric and magnetic multipoles.  Therefore, the observable physics
one can associate with toroidal moments in crystalline solids is
indistinguishable from the physics due to electric and magnetic
multipoles.

We illustrate our general theory taking polarized versions of
lonsdaleite and diamond as examples; two highly symmetric crystal
structures whose variations are realized in numerous technologically
relevant materials.
An overview of the specific types of crystal structures considered
in the present work is presented in Table~\ref{tab:symmetries}.  The
space-inversion and time-inversion symmetric pristine lonsdaleite
and diamond structures are compatible with even\=/$\ell$ electric
polarizations of order $\ell \ge 2$ for lonsdaleite and $\ell \ge 4$
for diamond, respectively \cite{kos63}.  Introduction of
odd\=/$\ell$ electric polarizations (also $\ell = 2$ in diamond) and
even- or odd\=/$\ell$ magnetic polarizations reduce the crystal
symmetries to those specified by the magnetic point groups
\cite{lan8e, lud96} given in Table~\ref{tab:symmetries}.  For
example, SIS of the lonsdaleite structure is broken in the wurtzite
structure with an $\ell=1$ electropolarization (i.e., an electric
dipolarization).  Similarly, the zincblende structure constitutes a
broken-SIS diamond structure due to an $\ell=3$ electropolarization
(i.e., an electric octupolarization).  We also consider the
structures where TIS is broken instead of SIS by introducing
different magnetic polarizations ranging from $\ell = 1$ to $\ell =
4$.  Lonsdaleite and diamond structures with a
magnetic quadrupolarization ($\ell = 2$) or hexadecapolarization
($\ell = 4$) are antiferromagnets \cite{oit18} that break both SIS
and TIS individually but preserve CIS.

The remainder of this article is organized as follows.  We start by
developing the general theory in Sec.~\ref{sec:theory}.  Results
obtained from application of the theory to crystal structures
from the lonsdaleite and diamond families are presented in the
subsequent Secs.~\ref{sec:lons} and \ref{sec:diamond}, respectively.
Each main section has a preamble that gives a more detailed overview
of the topics discussed there.  Conclusions and a brief outlook are
presented in Sec.~\ref{sec:conclusions}.

%%%%%%%%%%%%%%%%%%%%%%%%%%%%%%%%%%%%%%%%%%%%%%%%%%%%%%%%%%%%%%%%%%
\section{General theory}
\label{sec:theory}

In this section, we develop a rigorous theory of multipolar order in crystalline solids based on group theory. The symmetry properties
of electric and magnetic multipoles in free space are discussed in
Sec.~\ref{sec:multipoles}, with special focus on their classification
according to transformation behavior under space inversion $i$ and time
inversion $\theta$. Section~\ref{sec:categories} considers multipole
\emph{densities} in crystalline solids, which we refer to as
\emph{polarizations}. Five distinct  categories of polarized matter are
identified. Sections~\ref{sec:compat} and \ref{sec:theo-invariants} develop
the mathematical tools required to enable the derivation of distinctive
\emph{indicators} for multipolar order in Sec.~\ref{sec:indicators}. The
consideration of symmetry hierarchies in Sec.~\ref{sec:sym-hierarchies}
provides a platform for addressing the subtle issue of when and how
coexisting multipole densities can be properly defined in a solid.
Section~\ref{sec:categories:disp} identifies distinctive band-structure
features for each category of polarized matter. Toroidal moments are
discussed in Sec.~\ref{sec:toroid}. The final Sec.~\ref{sec:macromicro}
elucidates the relationship between macroscopic multipole densities in a
solid's bulk and microscopic multipoles localized on atomic sites.

\subsection{Multipoles}
\label{sec:multipoles}

We are interested in electric and magnetic multipolar order in a
crystalline environment.  In free space, electric and magnetic
multipoles $\vekc{M}$ of order $\ell$ can be viewed as spherical
tensors that transform irreducibly under the rotation group $\Rit$
\cite{misc:cartesian-tensors}, i.e., they can be classified by the
($2 \ell +1$)-dimensional IRs $D_\ell^{ss'}$ ($\ell = 0, 1, 2,
\ldots$) of $\Rit$ according to which these quantities transform
\cite{jac99, edm60, tun85}.  Below we distinguish the components of
these ($2 \ell +1$)-dimensional IRs via an index $m$ in the usual
way \cite{edm60, tun85}.

The \emph{signature} $ss'$ indicates how a quantity behaves under
space inversion (even/odd if $s = +/-$) and time inversion (even/odd
if $s' = +/-$).  For even $\ell$, electric (magnetic) multipoles
transform according to the IR $D_\ell^{++}$ ($D_\ell^{--}$) of
$\Rit$, while for odd $\ell$, electric (magnetic) multipoles
transform according to $D_\ell^{-+}$ ($D_\ell^{+-}$), see
Table~\ref{tab:tensor-signat}.  Thus even\=/$\ell$ electric
multipoles preserve both SIS and TIS, whereas even\=/$\ell$ magnetic
multipoles break both SIS and TIS, but the combined inversion $i
\theta$ remains a good symmetry.  Odd\=/$\ell$ electric (magnetic)
multipoles break SIS (TIS).  The distinct behavior of these
multipoles under SIS and TIS suggests to divide these multipoles
into four \emph{families} representing even\=/$\ell$ and
odd\=/$\ell$ electric and magnetic multipoles.

The behavior of the four families of electric and magnetic
multipoles under SIS and TIS can be classified via the five
\emph{inversion groups} $\gsistis$, $\gtis$, $\gsis$, $\gstis$ and
$C_1$ that can be formed from space inversion $i$ and time inversion
$\theta$.  (See the explicit definitions given in the left column of
Table~\ref{tab:mpole-even-odd}.)  Under the full inversion group
$\gsistis$, when $i$ and
$\theta$ are independently good symmetries, only even\=/$\ell$
electric multipoles (signature $++$) are allowed.  Under $\gtis$,
i.e., when $i$ is broken, we may also have odd\=/$\ell$ electric
multipoles ($-+$).  Under $\gsis$, on the other hand, i.e., when
$\theta$ is broken, we may have instead odd\=/$\ell$ magnetic
multipoles ($+-$).  When both $i$ and $\theta$ are broken, but $i
\theta$ remains a good symmetry so that we get the group $\gstis$,
we may have even\=/$\ell$ magnetic multipoles ($--$).  Even\=/$\ell$
and odd\=/$\ell$ electric and magnetic multipoles (all signatures
$ss'$) are allowed simultaneously if none of the operations $i$,
$\theta$, and $i\theta$ represent good symmetries, i.e., we get the
trivial group $C_1$ that only contains the identity $e$.

\subsection{Categories of polarized matter}
\label{sec:categories}

An extended crystal must be characterized in terms of multipole
\emph{densities} $\vekc{m}$ instead of multipoles $\vekc{M}$.  To
emphasize the conceptual difference between the quantities
$\vekc{m}$, which constitute a macroscopic property of the bulk
material, and the localized multipoles $\vekc{M}$ characterizing,
e.g., molecules \cite{jac99, raa05}, we refer to $\vekc{m}$ as a
\emph{polarization}.  Nonetheless, from the perspective of group
theory, both $\vekc{M}$ and $\vekc{m}$ are spherical tensors of
order $\ell$ that share the same transformation properties under the
point-group symmetries discussed here.  The multipoles $\vekc{M}$ of
localized systems such as molecules can be characterized in terms of
the point groups $G$ characterizing these systems \cite{her50}.
Similarly, in a crystalline environment, the symmetry is reduced
compared with free space.  According to Neumann's principle, the
relevant symmetry group for material tensors such as multipole
densities is the crystallographic point group $G$ defining the
crystal class of a crystal structure \cite{voi10, nye85, bir74}.
These groups are finite subgroups of the rotation group $\Rit$.

In total, 122 magnetic crystallographic point groups $G$ can be
formed \cite{tav56, cor58a, bra72}.  To categorize these groups, we
expand on the classification of multipolar order based on the
inversion symmetries $\gamma = i$, $\theta$, and $i\theta$.  We
decompose
\begin{equation}
  \label{eq:group-inv-decomp}
  G = \tilde{G} \times C_\gamma \,,
\end{equation}
where $\tilde{G}$ denotes the proper or improper subgroup of $G$
that contains none of the inversion symmetries $\gamma$ as individual
group elements, and $C_\gamma$ is the inversion group
that can be formed from the inversion symmetries $\gamma$ that
appear as group elements in $G$.  In this way we identify five
qualitatively distinct categories of \emph{macroscopic}
electromagnetic multipolar order based on the five inversion groups
$C_\gamma$ \cite{misc:incommensurate}.  The five categories are
listed in the last column of Table~\ref{tab:mpole-even-odd}, and
their properties are as follows.

(i) The full inversion group $\gsistis$ defines \emph{parapolar} systems that
have the highest symmetry.  Even\=/$\ell$ electric multipole densities
(signature $++$) are the only electromagnetic multipole densities
permitted by $\gsistis$, and we use the label \emph{parapolarization}
for members of this family.  Parapolar systems are thus both
paraelectric and paramagnetic.

(ii) The non-cyclic group $\gsistis$ has order 4.  It is isomorphic
to the Klein four-group in abstract group theory \cite{lud96}.
Accordingly, the group $\gsistis$ has three order\=/2 subgroups
$\gtis$, $\gsis$, and $\gstis$ that represent mutually exclusive
alternatives to reduce the symmetry of parapolar systems.  We
elaborate on each one of these in turn.  (ii\=/a)~The group $\gtis$
defines \emph{electropolar} systems that may possess odd\=/$\ell$
electric multipole densities (signature $-+$, any member of this family
is labeled an \emph{electropolarization}), including an electric dipole
density (an electric dipolarization with $\ell = 1$).  Therefore,
electropolar systems include, e.g., pyroelectrics and ferroelectrics
\cite{voi10, nye85, new05, tin08}.  (ii\=/b)~The group $\gsis$ defines
\emph{magnetopolar} systems that may possess odd\=/$\ell$ magnetic
multipole densities (a \emph{magnetopolarization}, $+-$), including a
magnetic dipole density (a magnetization with $\ell = 1$).  Therefore,
magnetopolar systems include, e.g., ferromagnets \cite{tav58, cor58a,
bir64, new05}.  (ii\=/c)~The group $\gstis$ defines
\emph{antimagnetopolar} systems that may possess even\=/$\ell$
magnetic multipole densities (an \emph{antimagnetopolarization},
$--$).  Antiferromagnets can be antimagnetopolar or magnetopolar as
illustrated below \cite{misc:iv-space-groups}.
We denote the electropolar, magnetopolar and antimagnetopolar groups
jointly as \emph{unipolar} groups.

(iii) The trivial inversion group $C_1$ defines
\emph{multipolar} systems that may possess electric and magnetic
multipole densities of any order $\ell$ (all signatures $ss'$) so that
they can be simultaneously electropolar and (anti\=/)magnetopolar.
Multipolar systems include, e.g., multiferroics \cite{fie16, spa17}.

The five inversion groups $C_\gamma$ treat the black-white
symmetries $i$ and $\theta$ on the same footing.  Closely related,
the five categories in Table~\ref{tab:mpole-even-odd} treat electric
and magnetic order on the same footing.  Here we complement the
standard Sch\"onflies notation for magnetic point groups
\cite{lud96, lan8e} with the corresponding expressions according to
Eq.\ (\ref{eq:group-inv-decomp}) that reveal how the inversion
operations are combined with proper rotations.

\subsection{Compatibility relations}
\label{sec:compat}

The crystallographic point groups $G$ are finite subgroups of the
rotation group $\Rit$, and the IRs $D_\ell^{ss'}$ of $\Rit$ can be
mapped onto the IRs $\Gamma_\alpha$ of $G$
\begin{equation}
  \label{eq:def:compat}
  \Rit \mapsto G : \quad D_\ell^{ss'} \mapsto \sum_\alpha \Gamma_\alpha \,.
\end{equation}
Ignoring TIS, i.e., considering only the nonmagnetic
crystallographic point groups, the \emph{compatibility relations}
(\ref{eq:def:compat}) were tabulated up to rank $\ell=6$ by Koster
\emph{et al.}\ \cite{kos63, misc:compat}.  These relations indicate
how spherical tensors $\vekc{m}$ decompose into components
$\vekc{m}^G_\alpha$ transforming irreducibly (IR $\Gamma_\alpha$)
under a crystallographic point group $G$
\begin{equation}
  \label{eq:multipole-spher-cryst}
  \vekc{m} = \sum_\alpha \vekc{m}^G_\alpha \,.
\end{equation}
For brevity of notation, we ignore in Eqs.\ (\ref{eq:def:compat})
and (\ref{eq:multipole-spher-cryst}) that there may be multiple
irreducible components $\vekc{m}^G_\alpha$ transforming according to
the same IR $\Gamma_\alpha$.  (This is expressed by the
multiplicities with which an IR $\Gamma_\alpha$ of $G$ is contained
in an IR $D_\ell^{ss'}$ of $\Rit$.)  The components
$\vekc{m}^G_\alpha$ can be obtained by projecting $\vekc{m}$ onto
the IRs $\Gamma_\alpha$ of $G$.  The IRs of the crystallographic
point groups are at most three-dimensional, i.e., spherical
multipole densities $\vekc{m}$ of order $\ell \ge 2$ decompose into
multiple irreducible components under all crystallographic point
groups $G$.

We make extensive use of Koster's tables \cite{kos63}, though occasionally
we need to deviate from Koster's conventions regarding the choice of
coordinate systems used to define the group elements and basis
functions.  Koster \emph{et al.}\ define the IRs for the 32
nonmagnetic crystallographic point groups $G$ via their characters
and representative basis functions.  Koster \emph{et al.}\ consider
only SIS but not TIS, and they use a single superscript $s = \pm$ to
denote IRs that are even or odd under SIS.  So-called type-III
magnetic point groups \cite{bra72} are isomorphic to nonmagnetic
point groups (provided we do not consider double groups
\cite{misc:double}, as appropriate for Neumann's principle
\cite{bir74}).  However, in order to identify representative basis
functions for the IRs as tabulated by Koster \emph{et al.}\
\cite{kos63}, we consider the IRs of the respective nonmagnetic
subgroups of the type-III magnetic groups.  Below we indicate these
homomorphisms relating the magnetic groups with their nonmagnetic
subgroups via an arrow, e.g., $\Rit \rightarrow R_i \equiv R \times
\gsis$.  The approach
followed here is thus similar to how TIS can be taken into account
for nonmagnetic systems, starting with the representations of point
groups ignoring TIS and subsequently incorporating the effect of TIS
\cite{bir74}.

If the symmetry of a ``parent'' crystalline environment is reduced
from a group $G$ to a subgroup $U$ of $G$, the IRs $\Gamma_\alpha$
of $G$ can likewise be mapped onto the IRs $\Gamma_\beta$ of $U$
\cite{misc:compat-nonunique}
\begin{equation}
  G \mapsto U : \quad
  \Gamma_\alpha \mapsto \sum_\beta \Gamma_\beta \,.
\end{equation}
The compatibility relations for the IRs of the crystallographic
point groups have also been tabulated by Koster \emph{et al.}\
\cite{kos63}.  They imply that, similar to Eq.\
(\ref{eq:multipole-spher-cryst}), components $\vekc{m}^G_\alpha$ of
a multipole density $\vekc{m}$ that transform irreducibly under $G$
can be decomposed into components $\vekc{m}^U_\beta$ that transform
irreducibly under $U \subset G$
\begin{equation}
  \label{eq:multipole-cryst-cryst}
  \vekc{m}^G_\alpha = \sum_\beta \vekc{m}^U_\beta \,.
\end{equation}

The decomposition of a (spherical or cartesian) material tensor
$\vek{T}$ into its irreducible components under a point group $G$
enables one to determine which components of $\vek{T}$ are allowed
to be nonzero according to crystal symmetry.  Components of a
material tensor $\vek{T}$ are allowed by crystal symmetry
\cite{bir74} if these components transform according to the identity
(unit) representation of $G$ (always denoted $\Gamma_1$ in Koster's
notation \cite{kos63}).  For higher-rank cartesian tensors with
multiple indices, additional constraints for nonzero tensor
components may arise from symmetry under permutation of indices.
Nonzero tensor components for a range of common material tensors
have been tabulated for the 32 nonmagnetic crystallographic point
groups in, e.g., Refs.\ \cite{nye85, new05, tin08}.  Material tensors
have been discussed for the magnetic crystallographic point groups
in, e.g., Refs.\ \cite{new05, tin08, bir64}.

The criterion for nonzero tensor components implies that more tensor
components become allowed to be nonzero if the symmetry of a system
is reduced, e.g., via external perturbations or due to a phase
transition \cite{lud96}.  This can be derived in detail from the
compatibility relations between the IRs of the crystallographic
point groups \cite{kos63}.  These techniques have been exploited
earlier to study and characterize material tensors \cite{nye85}.

The above considerations apply, in particular, to electric and
magnetic multipole densities $\vekc{m}$.  The decomposition
(\ref{eq:multipole-spher-cryst}) can be performed for multipole
densities of any order $\ell$ and any crystallographic point group
$G$ \cite{kos63}.  However, $\vekc{m}$ remains forbidden by symmetry
unless the decomposition (\ref{eq:multipole-spher-cryst}) includes
the identity representation $\Gamma_1$ of~$G$.

Phrased differently, suppose a crystal structure has a group $G$
that requires $\vekc{m} = \vek{0}$ for a given multipole density of
order $\ell$, so that Eq.\ (\ref{eq:multipole-spher-cryst}) does not
include a term associated with $\Gamma_1$.  If the multipole density
$\vekc{m}$ becomes nonzero (e.g., because of external perturbations
or due to a phase transition), the symmetry of the system is reduced
from $G$ to a subgroup $U$ of $G$ that is \emph{defined} by the
condition that the nonzero component of $\vekc{m}$ denoted
$\mathcal{m}^U_1$ transforms irreducibly according to the identity
representation $\Gamma_1$ of $U$, i.e., $\mathcal{m}^U_1$ is a
scalar under $U$ \cite{misc:struct-phase-trans}.  Again, we ignore
for brevity of notation that $\vekc{m}$ may contain multiple
distinct components that transform according to $\Gamma_1$ of $U$.
We call $\mathcal{m}^U_1$ the scalar of the multipole density
$\vekc{m}$ under~$U$.

The subgroup $U \subset G$ generally depends on which component of
$\vekc{m}$ has become nonzero.  For electric and magnetic dipole
densities, the resulting subgroups $U$ have likewise been tabulated
by Koster \emph{et al.}\ \cite{kos63}.  Groups permitting a
macroscopic electric dipole density ($\ell=1$) have previously been
called polar \cite{nye85}, and groups permitting a magnetic dipole
density (magnetization) represent ferromagnetism \cite{tav58, cor58a,
bir64, new05}.  According to Neumann's principle, the respective
point groups $G$ can be identified via the criterion that the
compatibility relation (\ref{eq:def:compat}) for the IR $D_1^{-+}$
($D_1^{+-}$) of the electric (magnetic) dipole density includes the
identity representation $\Gamma_1$ of $G$.  More explicitly, we
obtain the nonzero component $\mathcal{m}_1$ of $\vekc{m}$ by
projecting $\vekc{m}$ onto $\Gamma_1$ of $G$ \cite{bir74}.

More generally, we can classify the crystallographic point groups
$G$ based on the lowest-order electric
($\ell_\mathrm{min}^{(\mathrm{e},\lambda)}$) and magnetic
($\ell_\mathrm{min}^{(\mathrm{m},\lambda)}$) multipole densities
permitted in a crystal structure by its point group $G$ (ignoring
the electric monopole density transforming according to $D_0^{++}$
that is always allowed) \cite{kos63}.  The superscript $\lambda$
distinguishes between the lowest even ($\lambda = g$) and odd
($\lambda = u$) orders $\ell$ corresponding to the different
families of polarizations.  Generally, the higher the symmetry of a
system is (as characterized via its group $G$), the higher are the
orders $\ell_\mathrm{min}^{(\mathrm{e},\lambda)}$ and
$\ell_\mathrm{min}^{(\mathrm{m},\lambda)}$, with
$\ell_\mathrm{min}^{(\mathrm{e},u)} = \infty$ when SIS is a good
symmetry and $\ell_\mathrm{min}^{(\mathrm{m},\lambda)} = \infty$ for
nonmagnetic groups.  The quantities
$\ell_\mathrm{min}^{(\mathrm{e},\lambda)}$ and
$\ell_\mathrm{min}^{(\mathrm{m},\lambda)}$ thus define a physically
motivated hierarchy among the crystallographic point groups $G$.
This symmetry-based classification of macroscopic electric and
magnetic order is independent of how the order is realized
microscopically.  For example, as discussed in
Sec.~\ref{sec:diamond:e4}, an $sp^3$ tight-binding (TB) model can
describe macroscopic electric hexadecapole densities ($\ell = 4$), and
locally alternating magnetic dipoles can give rise to quadrupolar magnetic
order (Secs.~\ref{sec:lons:m2} and \ref{sec:diamond:m2}).

As per the definition of $\ell_\mathrm{min}^{(\mathrm{e},\lambda)}$ and
$\ell_\mathrm{min}^{(\mathrm{m},\lambda)}$, a group $G$ may also permit
electromagnetic multipole densities of higher order than
$\ell_\mathrm{min}^{(\mathrm{e},\lambda)}$ and
$\ell_\mathrm{min}^{(\mathrm{m},\lambda)}$.  However, it is
well-known for finite, localized systems that the higher-order
multipoles are generally not well-defined \cite{jac99, raa05}.  We
discuss this point in greater detail in the context of bulk multipole
densities of crystalline solids in Sec.~\ref{sec:sym-hierarchies}.

\subsection{Theory of invariants}
\label{sec:theo-invariants}

The theory of invariants \cite{lut56, bir74, win03} provides a
systematic framework to describe the dynamics of Bloch electrons in
the presence of perturbations such as electric and magnetic
multipole densities.  For conceptual clarity we restrict the
discussion in the present work to nondegenerate bands, though it is
well-known how the theory of invariants can be extended to bands
involving degeneracies \cite{bir74, win03}.  We consider a system
with crystallographic magnetic point group $G$.  The general
arguments presented in this section apply to any crystallographic
point group $G$.  Later on, we focus specifically on $G = D_{6h}
\times \gtis$ (lonsdaleite) and $G = O_h \times \gtis$ (diamond).  Note
that these two groups constitute the highest-symmetry crystallographic
point groups.  All nonmagnetic crystal structures have point groups that
are proper or improper subgroups of $D_{6h}$ or $O_h$ (ignoring TIS;
see Fig.~5 in Ref.\ \cite{kos63}), and we get subgroups of
$D_{6h}\times \gtis$ and $O_h \times \gtis$ for magnetic structures.

The theory of invariants is based on the fact that the Hamiltonian
must transform according to the identity representation $\Gamma_1$
of $G$.  More generally, \emph{any} operator transforming according
to $\Gamma_1$ of $G$ has a nonzero expectation value (unless the
system possesses ``hidden symmetries'' so that $G$ is not actually
the symmetry group of the system).  In contrast, the expectation
values of operators not transforming according to $\Gamma_1$ of $G$
must vanish.

In the theory of invariants, the Hamiltonian is built up from
invariants that transform each according to the identity
representation $\Gamma_1$ of $G$.  Such invariants can be expressed
in terms of scalar products of tensor operators $\vek{O}$ and
$\vek{O}'$ that transform according to complex-conjugate
representations $\Gamma$ and $\Gamma^\ast$ of $G$ (because the
product representation $\Gamma \times \Gamma^\ast$ necessarily
contains $\Gamma_1$).  It is preferable, though not necessary, that
$\vek{O}$ and $\vek{O}'$ are chosen such that their representations
$\Gamma$ and $\Gamma^\ast$ are irreducible under $G$.  For the
groups $D_{6h} \times \gtis$, $O_h \times \gtis$, and relevant
subgroups discussed below, all IRs are real, $\Gamma^\ast = \Gamma$.

We may add terms to the Hamiltonian that describe the effect of
electric and magnetic multipole densities $\vekc{m}$.  The
respective invariant interaction terms can be written as a sum of
scalar products \cite{bir74}
\begin{equation}
  \label{eq:invars}
  \sum_\alpha a^G_\alpha \, \vek{K}^G_\alpha \cdot \vekc{m}^G_\alpha \,,
\end{equation}
where the sum runs over the IRs $\Gamma_\alpha$ appearing in the
decomposition (\ref{eq:multipole-spher-cryst}), and the irreducible
tensor operators $\vek{K}^G_\alpha$ transform according to the
respective complex-conjugate IRs $\Gamma_\alpha^\ast$.  For brevity
of notation, we ignore that, in general, we have multiple
irreducible tensors $\vekc{m}^G_\alpha$ ($\vek{K}^G_\alpha$) that
transform according to the same IR $\Gamma_\alpha$
($\Gamma_\alpha^\ast$) of $G$.  Furthermore, we restrict the present
analysis to effects linear in the multipole densities $\vekc{m}$.
The irreducible tensors $\vekc{m}^G_\alpha$ can also be used to
construct irreducible tensors of higher degree in the components of
$\vekc{m}$ (similar to the tensor operators $\vek{K}^G_\alpha$
discussed below that may be higher-degree polynomials in the
components of crystal momentum $\hbar\vek{k}$).  The expansion
coefficients $a^G_\alpha$ are material-specific parameters that are
generally different for different electronic bands.

\begin{table}[b]
  \caption{\label{tab:tensor-op-power} Powers of cartesian components
  of the wave vector (collectively denoted $k$) and components of
  spin (collectively denoted $\sigma$) required for a polynomial
  representation of tensor operators with signature $ss'$ associated
  with multipoles of order $\ell$.  The symbol $n$ denotes a
  non-negative integer.}
  \renewcommand{\arraystretch}{1.2}
  \let\mc\multicolumn
  \tabcolsep 0pt
  \begin{tabular*}{\linewidth}{cE*{4}{Ce@{\;:\;\;}LE}}
    \hline \hline \rule{0pt}{3.8ex}
    & \mc{2}{c}{electric} & \mc{2}{c}{magnetic}
    & \mc{2}{c}{\marr{electro- \\ toroidal}}
    & \mc{2}{c}{\marr{magneto- \\ toroidal}} \\ \hline \rule{0pt}{2.8ex}
    $\ell$ even
    & ++ & k^{2n+2} & -- & k^{2n+1}
    & -+ & k^{2n+1} \sigma & +- & k^{2n} \sigma \\
    $\ell$ odd
    & -+ & k^{2n+1} \sigma & +- & k^{2n} \sigma
    & ++ & k^{2n+2} & -- & k^{2n+1} \\ \hline \hline
  \end{tabular*}
\end{table}

When the theory of invariants is applied to the dynamics of Bloch
electrons, the components of the tensor operators $\vek{K}^G_\alpha$
are polynomials in the components of crystal momentum $\hbar\vek{k}$
and of spin $(\hbar/2) \vek{\sigma}$ (and sometimes also components
of orbital angular momentum representing band degeneracies).  Given
the transformational behavior of $\kk$ and $\vek{\sigma}$ under $G$,
\emph{all} tensor operators $\vek{K}^G_\alpha$ of given degrees in
$\kk$ and $\vek{\sigma}$ [even beyond those relevant for the
decomposition (\ref{eq:multipole-spher-cryst})] can be derived in a
systematic way using the coupling coefficients tabulated by Koster
\emph{et al.}  \cite{kos63}.  This makes the theory of invariants a
comprehensive theory regarding the effects induced by multipolar
order on the dynamics of Bloch electrons.  For systems with
spherical symmetry and ignoring SIS and TIS so that the symmetry
group is $R$, the tensor operators must transform according
to the IR $D_\ell$ of the multipole density $\vekc{m}$, compare Eq.\
(\ref{eq:SpherHamCoupl}) below.  In this case, the components of the
tensor operators $\vek{K}^R$ can be chosen to be the familiar
harmonic polynomials, the degree of which equals the order $\ell$ of
$\vekc{m}$ \cite{edm60}.  When the symmetry group $G$ is a finite
subgroup of $\Rit$, the degree of the polynomials $\vek{K}^G_\alpha$
need not match the order $\ell$ of the corresponding multipole
density $\vekc{m}$.  Instead, SIS and TIS require that the signature
$ss'$ of $\vekc{m}$ must equal the signature of the tensor operators
$\vek{K}^G_\alpha$.  Therefore, the signatures of $\vek{k}$ ($--$)
and $\vek{\sigma}$ ($+-$) represent constraints on the degree of the
polynomials $\vek{K}^G_\alpha$, see Table~\ref{tab:tensor-op-power}.
(We do not consider higher powers in $\vek{\sigma}$ because
$\sigma_j^2 = \openone$.)  For example, tensor operators
$\vek{K}^G_\alpha$ associated with electric and magnetic multipole
densities for odd $\ell$ necessarily involve the spin operator
$(\hbar/2) \vek{\sigma}$.  Purely orbital operators
$\vek{K}^G_\alpha$ associated with odd $\ell$ may arise for
degenerate and off-diagonally coupled bands \cite{bir74, win03,
win20} that are not studied in the present work.  Concrete examples
for how tensor operators $\vek{K}^G_\alpha$ represent multipoles are
given in Secs.~\ref{sec:lons} and~\ref{sec:diamond}.

\subsection{Indicators of multipolar order}
\label{sec:indicators}

We begin with the case that, for a given group $G$, the invariant
expansion (\ref{eq:invars}) is formulated for a multipole density
$\vekc{m}$ of order $\ell < \ell_\mathrm{min}$ when $\vekc{m} =
\vek{0}$, and the sum over IRs $\Gamma_\alpha$ of $G$ does not
include the identity representation $\Gamma_1$.  In this case, the
expectation values of all irreducible tensor operators
$\vek{K}^G_\alpha$ appearing in the expansion (\ref{eq:invars})
must vanish, $\avg{\vek{K}^G_\alpha} = \vek{0}$.

As discussed above, if, starting from a group $G$ that requires
$\vekc{m} = \vek{0}$, the multipole density $\vekc{m}$ becomes
nonzero (e.g., because of external perturbations or due to a phase
transition), the symmetry of the system is reduced from $G$ to a
subgroup $U$ of $G$, and the nonzero component $\mathcal{m}^U_1$ of
$\vekc{m}$ transforms according to the identity representation
$\Gamma_1$ of $U$.  This implies, in turn, that the corresponding
tensor operator $K^U_1$ also transforms irreducibly according to
$\Gamma_1$ of $U$ and has a nonzero expectation value,
$\avg{K^U_1} \ne 0$.  The tensor operator $K^U_1$ thus provides a
probe for the presence of the multipole density $\mathcal{m}^U_1$.

Given a multipole density $\vekc{m}$ for some $\ell$, under $G$ each
irreducible component $\vekc{m}^G_\alpha$ of $\vekc{m}$ defines the
\emph{indicator}
\begin{equation}
  \label{eq:indicator}
  \vek{I}^G_\alpha
  = \frac{\partial H}{\partial \vekc{m}^G_\alpha}
  = a^G_\alpha \, \vek{K}^G_\alpha
\end{equation}
as an operator that is independent of the presence of the multipole
density $\vekc{m}$.  In a system without multipolar order, i.e., when
the group $G$ requires $\vekc{m} = \vek{0}$, the expectation value
$\avg{\vek{I}^G_\alpha}$ must vanish, $\avg{\vek{I}^G_\alpha}
= \vek{0}$ because $\Gamma_\alpha \ne \Gamma_1$.  Conversely, when
some component $\mathcal{m}^U_1$ of $\vekc{m}^G_\alpha$ becomes
finite, the respective component $\avg{I^U_1}$ of
$\avg{\vek{I}^G_\alpha}$ becomes nonzero.
For small $|\vekc{m}^G_\alpha|$, the expectation value
$\avg{\vek{I}^G_\alpha}$ is given by the linear-response
expression
\begin{equation}
  \label{eq:av-ind}
  \avg{\vek{I}^G_\alpha} = \chi^G_\alpha \, \vekc{m}^G_\alpha \,,
\end{equation}
with the matrix $\chi^G_\alpha$ denoting the static uniform
$\vek{I}^G_\alpha$-$\vek{I}^G_\alpha$ response function \cite{giu05}
in the parent structure with group $G$.  According to Eq.\
(\ref{eq:av-ind}), the expectation value $\avg{\vek{I}^G_\alpha}$ is
a direct quantitative probe of the multipole density $\vekc{m}$.
Therefore, in the system with point group $U$, where the nonzero
component of $\avg{\vek{I}^G_\alpha}$ is given by $\avg{I^U_1}$, we
have $\mathcal{m}^U_1 = \big[ \chi^G_\alpha \big]{}^{-1}
\avg{I^U_1}$.

A familiar example for the indicator formalism is given by exchange
coupling in ferromagnets.  Here the invariant is an exchange term
$(g/2) \, \bohrmag \, \vek{\sigma} \cdot \vekc{x}$ with magnetic
dipole density $\vekc{x}$.  The $g$-factor $g$ is generally an
effective parameter that characterizes the parent structure with
group~$G$; it may deviate from the free-electron value $g=2$
\cite{kje57, rot59}.  In an anisotropic crystal environment, the
exchange term may break up into multiple invariant terms
representing different crystallographic directions and weighted with
different $g$-factors $g$ \cite{bir74}.  The indicator representing
exchange coupling is the magnetic-moment operator $(g/2) \, \bohrmag
\, \vek{\sigma}$ whose nonzero expectation value signals the
presence of ferromagnetic order.  After the system has undergone a
phase transition to a ferromagnetic state, the nonzero component of
$\avg{(g/2)\, \bohrmag \, \vek{\sigma}}$ is given by
$\chi^G_\mathrm{P} \, \mathcal{x}^U_1$, with $\chi^G_\mathrm{P}$
denoting the Pauli susceptibility in the parent structure.  Knowing
the latter thus enables determination of the ferromagnetic
structure's magnetic dipole density via $\mathcal{x}^U_1 = \big[
\chi^G_\mathrm{P} \big]{}^{-1} \avg{(g/2)\, \bohrmag \,
\vek{\sigma}}$.

For a structure with group $U$ giving $\mathcal{m}^U_1 \ne 0$, the
parent structure with group $G \supsetneq U$ and $\vekc{m} = 0$
takes the role of a reference state.  The need for a reference state
also arises in the modern theories of electric dipolarization
\cite{kin93, res93} and orbital magnetization \cite{cer06, shi07}
that do not define these quantities on an absolute scale.  Instead,
they are defined as differences between two states of the material
that can be connected by an adiabatic switching process \cite{res10,
res20}.  Similarly, it was noticed in an early study of the
thermodynamics of pyroelectricity that only the differences between
the dipolarization in different states of the system are physically
significant \cite{voi10}.

\subsection{Symmetry hierarchies}
\label{sec:sym-hierarchies}

The irreducible components $\vekc{m}^G_\alpha$ and
$\vekc{m}^U_\beta$ of electric and magnetic multipole densities
$\vekc{m}$ of different order $\ell$ that are permitted by different
crystallographic groups $G$ and subgroups $U$ define a physically
motivated hierarchy among these groups.  This hierarchy is
complemented by a matching hierarchy of tensor operators
$\vek{K}^G_\alpha$ and $\vek{K}^U_\beta$ that obey the same sequence
of compatibility relations (for the complex conjugate IRs) as
$\vekc{m}^G_\alpha$, and $\vekc{m}^U_\beta$ [Eqs.\
(\ref{eq:multipole-spher-cryst}) and
(\ref{eq:multipole-cryst-cryst})].

The hierarchy can be extended to include the rotation group $\Rit$
at the top, known as spherical approximation to the dynamics of
Bloch electrons \cite{lip70, win03}.  Under $\Rit$, the invariant
interaction between a spherical multipole density $\vekc{m}$ and
crystal electrons can be written as a scalar product \cite{edm60}
similar to Eq.\ (\ref{eq:invars})
\begin{equation}
  \label{eq:SpherHamCoupl}
  a^R \vek{K}^R \cdot \vekc{m} \,.
\end{equation}
Here, conceptually similar to the tensor operators
$\vek{K}^G_\alpha$, the components of $\vek{K}^R$ are harmonic
polynomials in the components of crystal momentum $\hbar\vek{k}$ and
spin $(\hbar/2) \vek{\sigma}$.  Ignoring SIS and TIS, the harmonic polynomials
transforming according to the IR $D_\ell$ of $R$ can be chosen as
polynomials of degree $\ell$ \cite{edm60}.  With SIS and TIS taken
into account, the polynomials $\vek{K}^R$ must transform according
to the IR $D_\ell^{ss'}$ of $\Rit$ so that these polynomials are at least of
degree $\ell$ to be consistent with Table~\ref{tab:tensor-op-power}.  By
definition, an electric monopole density $\ell = 0$ transforms according to
the identity representation $D_0^{++}$ of $\Rit$, i.e., it is always allowed
in the spherical approximation.  The corresponding scalar tensor
operators $K^R$ are given by even powers of the wave vector
$\vek{k}$.

Often in applications of the theory of invariants to the dynamics of
Bloch electrons, the main physics is already captured by a
Hamiltonian $H$ with spherical symmetry, i.e., a Hamiltonian that is
invariant under the rotation group $\Rit$ \cite{lip70, win03}.
Starting from such a model we can add successively the effect of
electric and magnetic multipoles $\vekc{m}$ of decreasing order
$\ell$ till the symmetry is reduced to the actual symmetry group
$G_0$ of the real system.  (This concept has been called
\emph{symmetry hierarchy} \cite{win03} or \emph{hierarchy of
approximations} \cite{tre79}.  Conceptually, this approach is
closely related to the virtual crystal approximation employed in
first-principles electronic band-structure methods \cite{misc:vca}.)
The importance of different multipoles $\vekc{m}$ is reflected by
the magnitude of the prefactors $a^G$ appearing in the invariant
expansion for these multipoles.  Commonly, the magnitude of the
prefactors $a^G$ decreases with decreasing order $\ell$ of the
multipoles \cite{lip70, win03, suz74, tre79}.

Interestingly, this well-established scheme appears to violate the
fact known for finite localized systems that only the multipole
$\vekc{M}$ of lowest nonvanishing order $\ell = \ell_\mathrm{min}$
is well-defined because the coefficients of the higher-order
multipoles $\vekc{M}$ depend in general on the choice of origin
\cite{jac99, raa05}.  However, the situation is qualitatively
different in extended crystalline solids where even the multipole
density $\vekc{m}$ of lowest nonvanishing order $\ell =
\ell_\mathrm{min}$ cannot be defined in terms of multipole moments
per unit cell because this multipole moment depends on the arbitrary
definition of a unit cell \cite{mar74}.  The modern theory of
electric dipolarization and magnetization thus defines the $\ell=1$
multipole densities in terms of geometric phases that are
independent of the definition of the unit cell \cite{res94, res10}.

In crystalline solids, the non-uniqueness of higher-order multipoles
takes a distinct twist.  Within each family of polarizations (even\=/$\ell$
electric, odd\=/$\ell$ electric, even\=/$\ell$ magnetic, and odd\=/$\ell$
magnetic), the respective invariants appearing in the expansion
(\ref{eq:invars}) are not uniquely defined, as any linear combination of
invariants in a family is again an invariant in that family, see footnote~35
in Ref.\ \cite{suz74}.  For a crystal structure with symmetry group
$G_0$, this ambiguity is resolved, but only for the multipole
density $\vekc{m}$ of lowest nonvanishing order in each family, by
requiring that the tensor operator $K^{G_0}_1$ associated with
$\mathcal{m}^{G_0}_1$ has a vanishing projection on the identity
representation $\Gamma_1$ of the supergroup $G \supsetneq G_0$ that
characterizes the system when $\vekc{m} = \vek{0}$.  This implies that
even\=/$\ell$ electric (magnetic) multipole densities can be defined
(together with their associated invariants) independent of
odd\=/$\ell$ electric (magnetic) multipole densities.

For example, pristine diamond with point group $O_h$ supports an
electric hexadecapole density $\ell = 4$, see
Sec.~\ref{sec:diamond:e4}.  If the symmetry is reduced from $O_h = O
\times \gsis$ to $O$, the system also supports electric multipole
densities with odd $\ell$.  However, the lowest-order electric
multipole density with odd $\ell$ permitted by the group $O$ has
$\ell = 9$ (while the associated lowest-degree invariant is of Dirac
type, $a^\inv{e}{9} \mathcal{m}^\inv{e}{9} \, \vek{\sigma} \cdot
\vek{k}$, see Sec.~\ref{sec:conclusions}, i.e., it corresponds to
$n=0$ in the notation of Table~\ref{tab:tensor-op-power}).  For a
system with point group $O$, the electropolarization with $\ell=9$
and its associated invariant are well-defined, despite the
concurrent presence of a parapolarization with $\ell=4$
(hexadecapolarization).

\subsection{Band-dispersion characteristics of the categories of
polarized matter}
\label{sec:categories:disp}

\begin{table}[b]
  \caption{\label{tab:bands-degen} Band degeneracies imposed by
  inversion symmetries for the categories of polarized matter.
  Lines connect band energies that are equal.}
  \renewcommand{\arraystretch}{1.2}
  \let\mc\multicolumn
  \tabcolsep 0pt
  \unitlength 0.24ex
  \newcommand{\mybox}[1]{\makebox[8ex]{$#1$}}
  \begin{tabular*}{63ex}{lELcccc@{}}
    \hline \hline
    \rule{0pt}{2.5ex}%
    category &
    & \mybox{E_\sigma (\kk)}
    & \mybox{E_\sigma (-\kk)}
    & \mybox{E_{\bar{\sigma}} (\kk)}
    & \mybox{E_{\bar{\sigma}} (-\kk)} \\ \hline \rule{0pt}{2.5ex}%
    parapolar & \gsistis &
    \makebox[0pt][l]{\begin{picture}(123,2)
      \multiput(0,4)(41,0){4}{\circle*{4}}
      \put(0,4){\line(1,0){123}}
     \end{picture}} \\
    electropolar & \gtis &
    \makebox[0pt][l]{\begin{picture}(123,2)
      \multiput(0,4)(41,0){4}{\circle*{4}}
      \put(43,4){\line(1,0){41}}
      \put(0,-1){\line(1,0){123}}
      \multiput(0,-1)(123,0){2}{\line(0,1){5}}
     \end{picture}} \\
    magnetopolar & \gsis &
    \makebox[0pt][l]{\begin{picture}(123,2)
      \multiput(0,4)(41,0){4}{\circle*{4}}
      \multiput(0,4)(82,0){2}{\line(1,0){41}}
     \end{picture}} \\
    antimagnetopolar & \gstis &
    \makebox[0pt][l]{\begin{picture}(123,2)
      \multiput(0,3)(41,0){4}{\circle*{4}}
      \put(0,-2){\line(1,0){82}}
      \multiput(0,-2)(82,0){2}{\line(0,1){5}}
      \put(41,8){\line(1,0){82}}
      \multiput(41,3)(82,0){2}{\line(0,1){5}}
     \end{picture}} \\
    multipolar & C_1 &
    \makebox[0pt][l]{\begin{picture}(123,2)
      \multiput(0,2)(41,0){4}{\circle*{4}}
     \end{picture}} \\ \hline \hline
  \end{tabular*}
\end{table}

\begin{table*}[t]
  \caption{\label{tab:mult-op} Invariants associated with multipolar order in
   the diamond family. Here $\vekc{m}^{(\mathrm{e}, \ell)}$ and
   $\vekc{m}^{(\mathrm{m}, \ell)}$ denote the order-$\ell$ electric and
   magnetic multipole densities considered in Sec.~\ref{sec:diamond}, and
   $n$ is the non-negative integer defined in
   Table~\ref{tab:tensor-op-power}.}
  \renewcommand{\arraystretch}{1.2}
  \let\mc\multicolumn
  \newcommand{\muvek}{\mathcal{V}}
  \tabcolsep 0pt
  \begin{tabular*}{1.0\linewidth}{CELLL}
    \hline \hline \rule{0pt}{2.8ex}
    ss' & \mc{1}{C}{n=0} & \mc{1}{C}{n = 1} & \mc{1}{C}{n > 1}  \\ \hline \rule{0pt}{3.0ex}
    ++ & \mathcal{m}^\inv{e}{2}_x \, (- 2k_x^2 + k_y^2 + k_z^2) + \cp
    \, \footnotemark[1] \footnotetext{Realized in strained or quantum-confined
    diamond.}
    & \mathcal{m}^\inv{e}{4} (k_x^2 k_y^2 + \cp) \\ \hline \rule{0pt}{3.0ex}
    -+ & \mathcal{m}^\inv{e}{1}_x  \, (\sigma_y k_z - \sigma_z k_y) + \cp
    \, \footnotemark[1]
    & \mathcal{m}^\inv{e}{3} [\sigma_x k_x (k_y^2 - k_z^2) + \cp] \\
    +- & \mathcal{m}^\inv{m}{1}_x \, \sigma_x + \cp
    & \mathcal{m}^\inv{m}{3} (\sigma_x k_y k_z + \cp) \\
    -- & \mathcal{m}^\inv{m}{2}_x \, k_x + \cp \, \footnote{Realized in strained 
    or quantum-confined diamond antiferromagnets.}
    & \mathcal{m}^\inv{m}{2}_x \, k_x (k_y^2 - k_z^2) + \cp
    & \mathcal{m}^\inv{m}{4} k_x k_y k_z (k_y^2-k_z^2) (k_z^2-k_x^2) (k_x^2-k_y^2)
    \\ \hline\hline
  \end{tabular*}
\end{table*}

\begin{figure*}[t]
  \centering
  \includegraphics[width=0.85\linewidth]{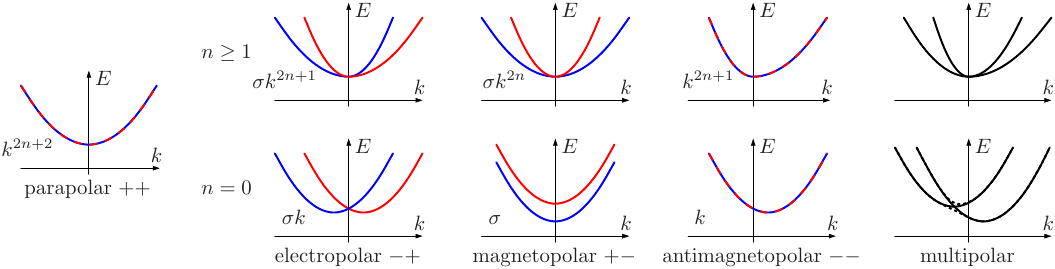}
  \caption{\label{fig:degen} Typical examples for spinful band
  dispersions $E_\sigma (\vek{k})$ associated with the five
  categories of polarized matter.  The bands for the parapolar and
  antimagnetopolar categories are at least twofold spin-degenerate
  in the entire Brillouin zone.  The upper (lower) row corresponds
  to $n \ge 1$ ($n = 0$), where $n$ is defined in
  Table~\ref{tab:tensor-op-power}.  The expressions in the lower left
  of the panels use the simplified notation of
  Table~\ref{tab:tensor-op-power} to represent the indicators for the
  presence of multipolar order.  Different colors represent opposite
  spin orientations.  In the multipolar case, the spin-split bands
  have more complicated spin textures such that it is generally not
  possible to assign a spin index to these bands.}
\end{figure*}

According to Table~\ref{tab:tensor-op-power}, the five categories of
polarized matter have unique patterns of band dispersions
$E_\sigma (\vek{k})$ as sketched in Fig.~\ref{fig:degen}.  Band degeneracies
for the five categories are summarized in
Table~\ref{tab:bands-degen}.  The parapolar category with the
highest symmetry is characterized by a spin-degenerate band
dispersion that only involves even powers of the wave vector
$\vek{k}$.  The patterns exhibited by the remaining four categories
depend on the non-negative integer $n$ determining the powers of the
cartesian components of the wave vector $\vek{k}$; see
Table~\ref{tab:tensor-op-power}.  Generally, the case $n=0$ is
qualitatively distinct from $n \ge 1$.  When $n=0$ (lower row in
Fig.~\ref{fig:degen}) the electropolar, the antimagnetopolar and the
multipolar categories are characterized via a finite slope of the
dispersion at $k = 0$.

More specifically, the patterns of band dispersions depend on the
particular tensor operators that couple to the multipole densities
$\vekc{m}$ according to the theory of invariants.  This is
illustrated in Table~\ref{tab:mult-op} for the invariants
associated with multipole densities in the diamond family.  (These
invariants are derived in Sec.~\ref{sec:diamond}.)  The electropolar
category with $n=0$ is realized by Rashba spin-orbit coupling
\cite{ras59a}, while the case $n=1$ includes Dresselhaus spin-orbit
coupling \cite{dre55}.  The magnetopolar category with $n=0$ yields
the exchange coupling in ferromagnets, while for $n=1$ it includes
altermagnets \cite{yua21, sme22, sme22a}.  An example for the
antimagnetopolar category with $n=1$ is the N\'eel term in diamond
antiferromagnets derived in Ref.\ \cite{win20}.

\subsection{Toroidal moments}
\label{sec:toroid}

For each $\ell = 0, 1, 2, \ldots$ the rotation group $\Rit$ has four
IRs $D_\ell^{ss'}$.  As discussed in Sec.~\ref{sec:multipoles}, for
even $\ell$, electric (magnetic) multipole densities transform
according to $D_\ell^{++}$ ($D_\ell^{--}$), while for odd $\ell$,
electric (magnetic) multipole densities transform according to
$D_\ell^{-+}$ ($D_\ell^{+-}$).  The remaining IRs for each $\ell$
have been associated with electrotoroidal ($D_\ell^{\mp +}$) and
magnetotoroidal ($D_\ell^{\pm -}$) moments \cite{art85, dub90,
spa08, pap16, nan16}, see Table~\ref{tab:tensor-signat}.  However,
the physical significance of toroidal moments has recently been
questioned~\cite{fer17}.

It was suggested that toroidal moments can be observed in
(magnetic) crystalline environments \cite{art85, dub90, spa08} when
these moments become symmetry-allowed, similar to the electric and
magnetic multipole densities discussed above \cite{misc:axions}.
However, unlike the rotation group $\Rit$, each of the (magnetic)
crystallographic point groups $G$ possesses only a finite set $\{
\Gamma_\alpha \}$ of IRs so that under symmetry reduction $\Rit
\mapsto G$, the toroidal moments are mapped onto the same set $\{
\Gamma_\alpha \}$ as the electric and magnetic moments \cite{kos63}.
According to Table~\ref{tab:mpole-even-odd}, each family of
electromagnetic multipoles with even (odd) $\ell$ has a matching
family of toroidal moments with odd (even) $\ell$, but the same
signature $ss'$.  These pairs of families thus have the same
transformational behavior under SIS and TIS and they couple to the
same indicators (Table~\ref{tab:tensor-op-power}).  Therefore, under
the smaller groups $G$, toroidal moments represent the same
observable physics as electric and magnetic moments.  Below, we
illustrate this point for crystal structures that are members of the
lonsdaleite and diamond families.  Toroidal moments are
fundamentally distinct from electromagnetic multipoles only under
the rotation group $\Rit$, but not under finite subgroups of $\Rit$.

Similar to electric and magnetic multipole densities $\vekc{m}$, for
each crystallographic group $G$ we can identify the lowest-order
electrotoroidal ($\ell_\mathrm{min}^{(\mathrm{et},\lambda)}$) and
magnetotoroidal ($\ell_\mathrm{min}^{(\mathrm{mt},\lambda)}$)
moments that are permitted by symmetry.
Generally there is no simple relation between, on the one hand,
$\ell_\mathrm{min}^{(\mathrm{e},\lambda)}$ and
$\ell_\mathrm{min}^{(\mathrm{m},\lambda)}$, and, on the other hand,
$\ell_\mathrm{min}^{(\mathrm{et}, \bar{\lambda})}$ and
$\ell_\mathrm{min}^{(\mathrm{mt}, \bar{\lambda})}$, where
$\bar{\lambda} = u$ ($\bar{\lambda} = g$) for $\lambda = g$
($\lambda = u$).  For example, the point group $D_{6h}$ of
lonsdaleite permits an electric quadrupole density ($\ell = 2$),
whereas the lowest electrotoroidal multipole permitted by $D_{6h}$
has $\ell = 7$.  Conversely, the lowest order of a magnetic
multipole density permitted by the group $D_{6h} (C_{6v})$ has $\ell
= 6$ whereas the same group permits a magnetotoroidal multipole with
$\ell = 1$.  This is similar to the fact discussed above that, under
crystallographic point groups $G$, the order $\ell$ of a multipole
density $\vekc{m}$ need not equal the polynomial degree of the
tensor operators $\vek{K}^G_\alpha$ associated with $\vekc{m}$ in
the invariant expansion (\ref{eq:invars}).

\subsection{Macroscopic multipole densities and localized
multipoles}
\label{sec:macromicro}

According to Neumann's principle, the pattern of macroscopic
multipole densities $\vekc{m}$ permitted in a crystal structure is
determined by the crystallographic point group $G$ defining the
crystal class of the crystal structure \cite{voi10, nye85, bir74}.
For crystal structures transforming according to a symmorphic space
groups $S$, the point group $G$ is the finite subgroup of $S$
consisting of the elements of $S$ that leave one point in space
fixed.  Nonsymmorphic space groups $S$ also contain group elements
that combine point group symmetries $g$ with nonprimitive
translations.  Here the elements $g$ are also elements of the point
group $G$, although these symmetry operations are not, by
themselves, elements of $S$.  The latter case makes crystallographic
point groups defining crystal classes qualitatively distinct from
point group symmetries of finite systems like molecules.

The macroscopic multipole densities $\vekc{m}$ can be realized
microscopically by localized multipoles $\vekc{M}$ that are arranged
periodically consistent with the space group $S$.  The length scale
of the localized multipoles $\vekc{M}$ is generally a fraction of
the lattice constant.  The permitted patterns of multipoles
$\vekc{M}$ are determined by the site symmetries characterizing the
Wyckoff positions of the atoms forming a crystal structure.  The
site symmetries are subgroups of the crystallographic point group
$G$; this is the reason why the order of the multipoles $\vekc{M}$
may be smaller (or larger) than the order of the resulting multipole
density $\vekc{m}$.  Generally, there is no simple relation between
the order of local multipoles $\vekc{M}$ and the order of the
resulting macroscopic multipole density $\vekc{m}$.  The site
symmetries are tabulated in Refs.\ \cite{hah05, lit13}, see also
Refs.\ \cite{sto05a, eva97}.

For example, the site symmetry of the atoms in the lonsdaleite
structure discussed below is the group $C_{3v}$ that permits a local
electric dipole moment, but also an electric octupole moment
\cite{kos63}.  The latter is realized by the $sp^3$ hybrid orbitals
with which elements like carbon form lonsdaleite.  A TB picture is
well-suited to discuss the local electronic structure of the atoms
as a function of their positions in a crystal structure, see
Ref.\ \cite{faj19} for a more detailed analysis.  The concept of site
symmetries and the localized multipoles permitted by these site
symmetries is independent of the ambiguous definition of a unit cell
for a crystal structure \cite{mar74}.

For magnetic structures, it is well-known that local magnetic dipole
moments on the atoms affect their site symmetries \cite{bur13}.
These constraints determine the magnetic space groups of these
structures and thus, in turn, also the magnetic point group $G$.  We
extend this scheme by also considering local magnetic multipoles of
higher order beyond $\ell = 1$ \cite{kur08}.  Throughout we consider
macroscopic electric and magnetic multipole densities on the same
footing.  Likewise, we consider local electric and magnetic
multipoles on the same footing.  The local multipoles $\vekc{M}$
attached to atomic sites in a crystal provide an instructive
physical picture for the microscopic origin of the macroscopic
multipole densities $\vekc{m}$.  Also, magnetic multipoles
$\vekc{M}$ attached to atomic sites provide a convenient means to
incorporate magnetic order into TB models.  For conceptual clarity,
we limit our discussion of examples in Secs.~\ref{sec:lons} and
\ref{sec:diamond} to configurations with local multipoles
$\vekc{M}$ of only one order $\ell$, even though computational
studies of real materials typically observe a greater variety of
such multipoles~\cite{spa13}.

\section{Lonsdaleite family}
\label{sec:lons}

As an illustration for how to apply the general theory developed in
Sec.~\ref{sec:theory}, we analyze electric and magnetic order in
variants of hexagonal lonsdaleite listed in
Table~\ref{tab:symmetries}.  In Sec.~\ref{sec:lons:e2}, we consider
the electric quadrupolarization that is already present in pristine
lonsdaleite and therefore exists alongside all other polarizations
that reduce the high symmetry of the lonsdaleite crystal structure.
We then discuss electropolarizations in Secs.~\ref{sec:lons:e3}
(electric octupolarization) and \ref{sec:lons:e1} (electric
dipolarization).  Magnetopolarizations are covered in
Secs.~\ref{sec:lons:m3} (magnetic octupolarization) and
\ref{sec:lons:m1} (magnetization), while results for
antimagnetopolarizations are presented in Secs.~\ref{sec:lons:m4}
(magnetic hexadecapolarization) and \ref{sec:lons:m2} (magnetic
quadrupolarization).  Section~\ref{sec:lons:e-m} is devoted to an
elucidation of close connections between electric and magnetic
orders.

To simplify the presentation, we always ignore TIS when analyzing
electric order.  As explained in Sec.~\ref{sec:compat}, we identify
IRs of magnetic point groups by referring to the respective
nonmagnetic subgroups as tabulated by Koster \emph{et al.}\
\cite{kos63}; and we indicate the homomorphisms relating the
magnetic groups with their nonmagnetic subgroups via an arrow
`$\rightarrow$'.  Within each subsection, we follow the same
outline.  We start by discussing the crystal symmetry and stating the
compatibility relations for the relevant multipole density.  We then
identify the terms in the Bloch-electron Hamiltonian linear in the
considered multipole density having lowest order in $\kk$ and
discuss associated physical ramifications.  Then we link the local
site symmetry with its allowed local multipoles to the bulk
multipole density.  We also identify the toroidal moment density
that manifests itself via the same invariants as the discussed
electric or magnetic multipole density.

\begin{figure*}[t]
  \centering
  \includegraphics[width=0.90\linewidth]{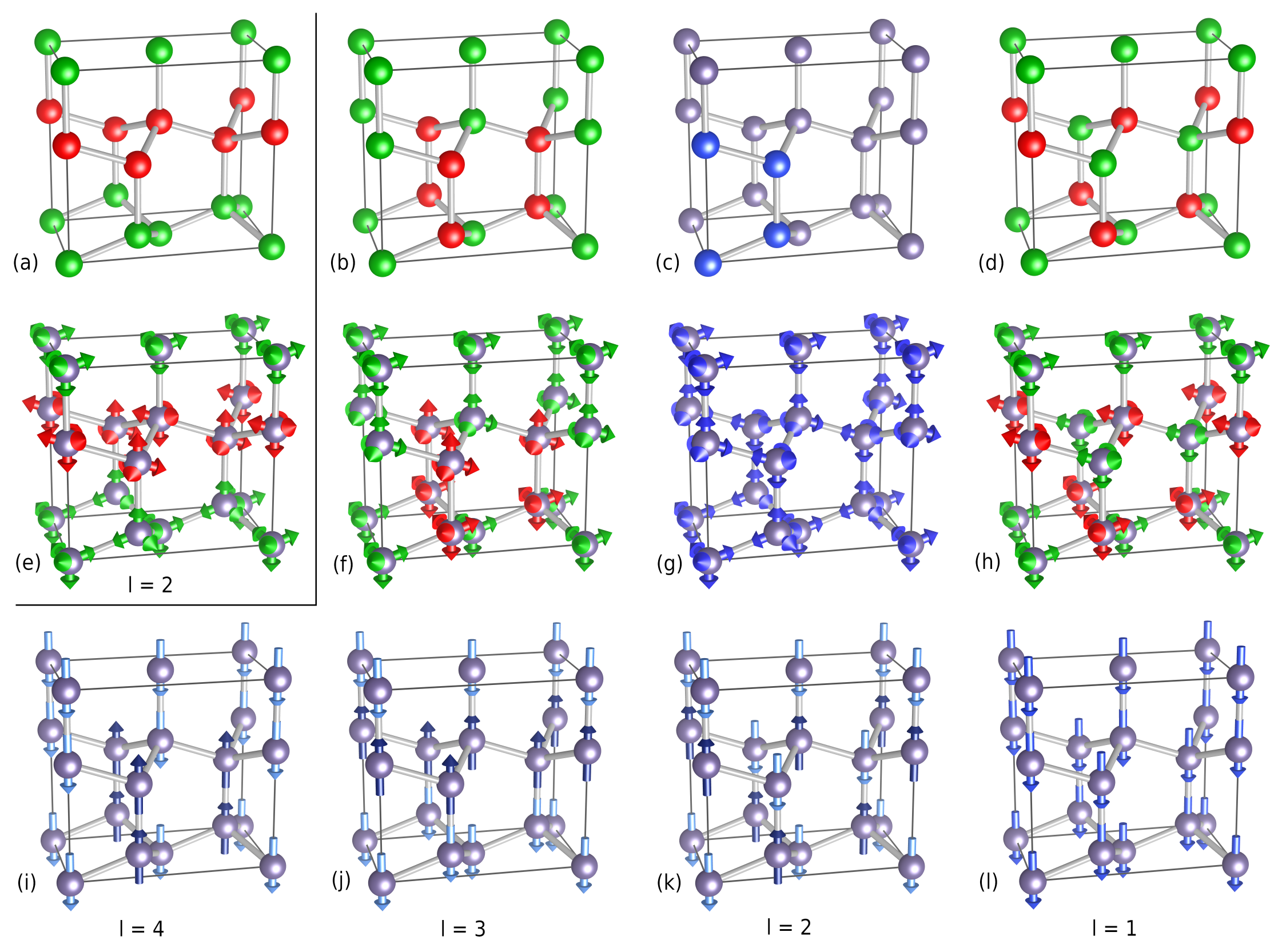}
  \caption{\label{fig:lons} Multipole densities in the lonsdaleite
  family.  Top row [(a), (b), (c), and (d)]: pristine lonsdaleite
  (c) and variants of the lonsdaleite crystal structure including
  wurtzite (d).  In (c), the four atoms in a unit cell are
  highlighted in blue.  (g) The local electric octupole moments
  $\vekc{M}_0$ on the identical atoms in pristine lonsdaleite give
  rise to an electric quadrupolarization ($\ell = 2$).  Remaining
  panels in the central row [(e), (f), and (h)]: For the structures
  (a), (b), and (d) consisting of two distinct types of atoms,
  panels (e), (f), and (h) show the deviation $\Delta\vekc{M}$ of
  the local octupole moments compared with the moments $\vekc{M}_0$
  when all atoms are identical [panel (g)].  These local moments
  $\Delta\vekc{M}$ give rise to (e) an electric quadrupolarization
  ($\ell = 2$), (f) an octupolarization ($\ell = 3$), and (h) a
  dipolarization ($\ell = 1$).  Bottom row [(i), (j), (k), and (l)]:
  local magnetic dipole moments give rise to (i) a
  hexadecapolarization ($\ell = 4$), (j) an octupolarization ($\ell
  = 3$), (k) a quadrupolarization ($\ell = 2$), and (l) a
  magnetization ($\ell = 1$).  Panels (i), (j), (k), and (l) show
  the local magnetic dipole moments with different shades of the
  same color because all sites are equivalent by symmetry (i.e.,
  they have the same Wyckoff letter) so that the local moments are
  likewise symmetry-equivalent.  The same situation arises for the
  electric octupole moments (g) of the nonmagnetic pristine
  lonsdaleite structure (c).}
\end{figure*}

\subsection{Electric quadrupolarization in pristine lonsdaleite}
\label{sec:lons:e2}

The hexagonal nonsymmorphic lonsdaleite structure is shown in
Fig.~\ref{fig:lons}(c).  The space group of lonsdaleite is $D_{6h}^4$
(No.~194, $P6_3/mmc$).  Ignoring TIS, the crystallographic point
group of lonsdaleite is $D_{6h} = D_6 \times \gsis$, i.e., lonsdaleite is
parapolar.  The lowest nonvanishing electric multipole density allowed
by $D_{6h}$ is an electric quadrupolarization ($\ell = 2$) with compatibility
relation \cite{kos63}
\begin{equation}
  R_i \mapsto D_{6h} : \quad
  D_2^+ \mapsto \Gamma_1^+ \oplus \Gamma_5^+ \oplus \Gamma_6^+ \,.
\end{equation}
Specifically, it is the component $m=0$ of the IR $D_2^+$ of
$R_i$ that transforms according to $\Gamma_1^+$ of $D_{6h}$ when the
symmetry is reduced from $R_i$ to $D_{6h}$.  Thus, written in
cartesian coordinates, the traceless electric quadrupole density in
lonsdaleite has nonzero components \cite{edm60, jac99}
\begin{equation}
  \label{eq:lons:e2:m}
  \mathcal{m}^\inv{e}{2} \equiv \mathcal{m}^\inv{e}{2}_{xx}
  = \mathcal{m}^\inv{e}{2}_{yy} = - \frack{1}{2} \,
  \mathcal{m}^\inv{e}{2}_{zz} \,.
\end{equation}
Here $\mathcal{m}^\inv{e}{2}$ is the scalar of the electric
quadrupole density under $D_{6h}$.  Accordingly, the harmonic
polynomial in the components of $\vek{k}$ for $\ell = 2$ and $m=0$
\begin{equation}
  K^\inv{e}{2}_1 = k_x^2 + k_y^2 - 2 \, k_z^2
\end{equation}
yields the invariant
\begin{align}
  \label{eq:lons:e2:H}
  H^\inv{e}{2}
  & = a^\inv{e}{2} \mathcal{m}^\inv{e}{2} K^\inv{e}{2}_1 \nonumber \\*
  & = a^\inv{e}{2} \mathcal{m}^\inv{e}{2} (k_x^2 + k_y^2 - 2 \, k_z^2) \,,
\end{align}
that represents an effective-mass anisotropy in the energy
dispersion $E_\sigma (\vek{k})$ of band electrons in lonsdaleite
\cite{ras59}.  The indicator $I^\inv{e}{2}_1 = a^\inv{e}{2} \,
K^\inv{e}{2}_1$ thus signals the electric quadrupole density in
lonsdaleite~\cite{misc:lons-indicator}.

The site symmetry of the atoms forming the lonsdaleite structure is
the group $C_{3v}$ \cite{hah05} that permits a local electric dipole
moment oriented along the main axis of lonsdaleite \cite{kos63}.
The group $C_{3v}$ also permits an electric octupole moment that is
naturally realized by the $sp^3$ hybrid orbitals with which elements
like carbon form lonsdaleite [Fig.~\ref{fig:lons}(g)], analogous to the
electric octupole moment of tetrahedrally bonded molecules such as
CH$_4$ \cite{sin64}.  It is the microscopic octupole that creates the
bulk quadrupolarization.  The electric quadrupolarization
(\ref{eq:lons:e2:m}) in pristine lonsdaleite and the associated
invariant (\ref{eq:lons:e2:H}) are also present in all other members
of the lonsdaleite family discussed in the remainder of this
section.  In that sense, the presence of the electric
quadrupolarization (\ref{eq:lons:e2:m}) represents a key feature of
the lonsdaleite family.

The lowest electrotoroidal moment permitted in pristine lonsdaleite
has $\ell = 7$.  It manifests itself via the same invariant
(\ref{eq:lons:e2:H}) as the electric quadrupole density.

\subsection{Electric octupolarization in lonsdaleite}
\label{sec:lons:e3}

Pristine lonsdaleite has four identical atoms per unit cell
[Fig.~\ref{fig:lons}(c)], and its lowest-order multipole density is
a quadrupolarization (Sec.~\ref{sec:lons:e2}).  Two distinct types
of atoms arranged as in Fig.~\ref{fig:lons}(b) reduce the
space-group symmetry to $D_{3h}^1$ (No.~187, $P\bar{6}m2$) and the
point-group symmetry is reduced to $D_{3h}$.  The group $D_{3h}$
breaks SIS so that this structure is electropolar.  The lowest-order
electric multipole density supported by the crystal structure in
Fig.~\ref{fig:lons}(b) is an electric octupole density ($\ell = 3$)
that yields the compatibility relation \cite{kos63}
\begin{equation}
  R_i \mapsto D_{3h} : \quad
  D_3^- \mapsto \Gamma_1 \oplus \Gamma_2 \oplus \Gamma_4
  \oplus \Gamma_5 \oplus \Gamma_6 \,.
\end{equation}
The corresponding invariants in the
Hamiltonian read (to lowest order in~$\vek{k}$)
\begin{align}
  \label{eq:lons:e3:H}
  H^\inv{e}{3} & = a^\inv{e}{3} \mathcal{m}^\inv{e}{3} \, k_z
  \left[ \sigma_x k_x k_y + \frack{1}{2} \sigma_y (k_x^2 - k_y^2) \right]
  \nonumber \\* & \hspace{1em} {}
  + b^\inv{e}{3} \mathcal{m}^\inv{e}{3} \, \sigma_z \, k_y
  \left( 3 k_x^2 - k_y^2 \right) \,.
\end{align}
These terms represent a spin-orbit coupling.

The site symmetry of the atoms is, once again, $C_{3v}$
\cite{hah05}.  However, the two distinct atoms in
Fig.~\ref{fig:lons}(b) have different Wyckoff positions (with two
atoms of each type per unit cell), and they may carry different
electric octupole moments.  We decompose these moments $\vekc{M}
\equiv \vekc{M}_0 + \Delta \vekc{M}$ into one part $\vekc{M}_0$ that
is equal by symmetry for all atoms as in lonsdaleite.  These
moments thus have the same observable effect as the local moments in
pristine lonsdaleite [Fig.~\ref{fig:lons}(g)], i.e., they give rise
to the invariant (\ref{eq:lons:e2:H}).  It is the remaining part
$\Delta \vekc{M}$ oriented oppositely that is shown in
Fig.~\ref{fig:lons}(f) and that gives rise to the new invariants
(\ref{eq:lons:e3:H}).  This can be worked out quantitatively in a
simple $sp^3$ TB model \cite{kob83}.

The lowest electrotoroidal moment permitted in this structure has
$\ell = 4$.

\subsection{Electric dipolarization in lonsdaleite -- wurtzite}
\label{sec:lons:e1}

If two of the four identical atoms in the lonsdaleite unit cell become
distinct as shown in Fig.~\ref{fig:lons}(d), we obtain the wurtzite structure
that is realized by several III\=/V and II\=/VI semiconductors
including ZnS, CdSe, GaN, and AlN.  The space group becomes
$C_{6v}^4$ (No.~186, $P6_3mc$), and the point group of wurtzite is
$C_{6v}$.  These groups break SIS so that wurtzite is electropolar.
More specifically, the compatibility relation \cite{kos63}
\begin{equation}
  R_i \mapsto C_{6v} : \quad
  D_1^- \mapsto \Gamma_1 \oplus \Gamma_5
\end{equation}
indicates that wurtzite naturally permits an electric dipole density
($\ell = 1$, an electric dipolarization).  The associated invariant in the
Hamiltonian to lowest order in $\kk$,
\begin{equation}
  \label{eq:lons:e1:H}
  H^\inv{e}{1} = a^\inv{e}{1} \mathcal{m}^\inv{e}{1}
  \left( \sigma_x k_y - \sigma_y k_x \right) \,\, ,
\end{equation}
represents a spin-orbit coupling commonly known as Rashba term
\cite{ras59a}.

A more complete analysis of the electric dipole density in wurtzite
is given in Table~\ref{tab:lons:e1}.  Ignoring TIS, the dipole
density transforms according to $D_1^-$ of $R_i$.  A hexagonal
environment (point group $D_{6h}$) yields the compatibility relation
\cite{kos63}
\begin{equation}
  R_i \mapsto D_{6h} : \quad
  D_1^- \mapsto \Gamma_2^- \oplus \Gamma_5^- \,,
\end{equation}
so that, as to be expected, a dipole density is forbidden in
lonsdaleite.  The lowest-order tensor operator transforming like
$\Gamma_2^-$ is
\begin{equation}
  \label{eq:lons:e1:K}
  K^\inv{e}{1}_{2-} = \sigma_x k_y - \sigma_y k_x \,,
\end{equation}
while two pairs of operators transform like $\Gamma_5^-$
\begin{equation}
    K^\inv{e}{1}_{5-} : \quad
    \sigma_y k_z, - \sigma_x k_z; \;
    \sigma_z k_y, - \sigma_z k_x \,.
\end{equation}
The expectation value of these operators must thus vanish in
lonsdaleite.  When the symmetry is further reduced to $C_{6v}$
(wurtzite), we get the compatibility relations \cite{kos63}
\begin{equation}
  \arraycolsep 0.1em
  D_{6h} \mapsto C_{6v} : \quad \left\{ 
  \begin{array}{rcls{0.6em}rcl}
    \Gamma_2^- & \mapsto & \Gamma_1, &
    \Gamma_5^- & \mapsto & \Gamma_5 \\[0.6ex]
    K^\inv{e}{1}_{2-} & \mapsto & K^\inv{e}{1}_1, &
    K^\inv{e}{1}_{5-} & \mapsto & K^\inv{e}{1}_5 \,,
  \end{array}
  \right.
\end{equation}
so that the tensor operator (\ref{eq:lons:e1:K}) becomes allowed and
yields the Rashba term (\ref{eq:lons:e1:H}).

Just as in Sec.~\ref{sec:lons:e3}, the distinct atoms in wurtzite
have different Wyckoff positions, but they all have the site
symmetry $C_{3v}$ \cite{hah05}.  The local electric moments on the
atoms are illustrated in Fig.~\ref{fig:lons}(h).

The lowest electrotoroidal moment permitted in wurtzite has $\ell = 6$.

\begin{table}
  \caption{\label{tab:lons:e1} \label{tab:lons:m1} Irreducible
  representations (IRs) and their lowest-order representative basis
  functions of an electric dipole (signature $-+$) and magnetic
  dipole (signature $+-$) oriented along the main axis of a
  hexagonal crystalline environment.  The IRs are labeled according
  to Koster \textit{et al.}\ \cite{kos63}.}
  \renewcommand{\arraystretch}{1.2}
  \let\mc\multicolumn
  \begin{tabular*}{\linewidth}{Ls{1em}ECCC}
    \hline \hline \rule{0pt}{2.8ex}
    \Rit \rightarrow R_i &
    \mc{3}{C}{D_1^{-+} \rightarrow D_1^-} \\[0.5ex] \hline \rule{0pt}{2.8ex}
    D_{6h} \times \gtis \rightarrow D_{6h} & \Gamma_2^- & \oplus & \Gamma_5^- \\
    & \sigma_x k_y - \sigma_y k_x &
    & \sigma_y k_z, - \sigma_x k_z; \\
    & & & \sigma_z k_y, - \sigma_z k_x \\
    C_{6v} \times \gtis \rightarrow C_{6v} &
    \Gamma_1 & & \Gamma_5 \\ \hline \rule{0pt}{2.8ex}
    \Rit \rightarrow R_i &
    \mc{3}{C}{D_1^{+-} \rightarrow D_1^+} \\[0.5ex] \hline \rule{0pt}{2.8ex}
    D_{6h} \times \gtis \rightarrow D_{6h} & \Gamma_2^+ & \oplus & \Gamma_5^+ \\
    & \sigma_z & & \sigma_x, \sigma_y \\
    D_{6h} (C_{6h}) \rightarrow C_{6h} &
    \Gamma_1^+ & & \Gamma_5^+ \oplus \Gamma_6^+ \\
    \hline \hline
  \end{tabular*}
\end{table}

\subsection{Magnetic hexadecapolarization in lonsdaleite}
\label{sec:lons:m4}

If the four equivalent atoms in the lonsdaleite unit cell carry
magnetic dipole moments as in Fig.~\ref{fig:lons}(i), the system
becomes antimagnetopolar.  More specifically, the magnetic space
group becomes $P6_3'/mm'c$ (No.~194.266) and the magnetic point
group becomes $D_{6h} (D_{3h}) \rightarrow D_{3h}$ that supports in
lowest order a magnetic hexadecapole density ($\ell = 4$) for which
the compatibility relation reads
\begin{equation}
  R_i \mapsto D_{3h} : \quad
  D_4^- \mapsto \Gamma_1 \oplus \Gamma_2 \oplus \Gamma_3
  \oplus 2 \Gamma_5 \oplus \Gamma_6 \,.
\end{equation}
The corresponding invariant in the
Hamiltonian reads (to lowest order in~$\vek{k}$)
\begin{equation}
  H^\inv{m}{4} = a^\inv{m}{4} \mathcal{m}^\inv{m}{4} \, k_x
  \left(k_x^2 - 3 k_y^2 \right) \,.
\end{equation}

The atoms with magnetic dipole moment aligned along the lonsdaleite
main axis have the site symmetry $C_{3v} (C_3)$.  The same site
symmetry is also realized in the magnetized versions of lonsdaleite
discussed in the remainder of this section.  In an $sp^3$ TB model
\cite{kob83} the magnetic dipoles can be implemented via a local
Zeeman term.

\subsection{Magnetic octupolarization in lonsdaleite}
\label{sec:lons:m3}

If the four equivalent atoms in the lonsdaleite unit cell carry
magnetic dipole moments as in Fig.~\ref{fig:lons}(j), the system
becomes magnetopolar.  More specifically, the magnetic space group
becomes $P6_3'/m'm'c$ (No.~194.268) and the magnetic point group
becomes $D_{6h} (D_{3d}) \rightarrow D_{3d}$ that supports in lowest
order a magnetic octupole density ($\ell = 3$) for which the
compatibility relation reads
\begin{equation}
  R_i \mapsto D_{3d} : \quad
  D_3^+ \mapsto \Gamma_1^+ \oplus 2 \Gamma_2^+ \oplus 2 \Gamma_3^+ \,.
\end{equation}
The corresponding invariant in the Hamiltonian reads (to lowest
order in~$\vek{k}$)
\begin{equation}
  \label{eq:lons:m3:Hxc}
  H^\inv{m}{3}_1 = a^\inv{m}{3} \mathcal{m}^\inv{m}{3} \,
  \sigma_z k_x k_z \left( k_x^2 - 3 k_y^2 \right) \,.
\end{equation}
Such a spin-splitting term induced by exchange coupling and
proportional to an even power of components of $\vek{k}$ has
recently been associated with altermagnetism \cite{sme22, sme22a}.
This term aligns the magnetic moments of the Bloch electrons
(anti-)parallel to the local magnetic moments on the individual atoms
[Fig.~\ref{fig:lons}(j)].  The combined effect of nonrelativistic
exchange coupling and relativistic spin-orbit coupling gives rise to
additional invariants
\begin{align}
  H^\inv{m}{3}_2 & = b^\inv{m}{3} \mathcal{m}^\inv{m}{3} \,
  \left( \sigma_x k_y - \sigma_y k_x \right) k_z
  \nonumber \\* & \hspace{1em} {}
  + c^\inv{m}{3} \mathcal{m}^\inv{m}{3}
  \left[ \frack{1}{2} \sigma_x (k_y^2 - k_x^2) + \sigma_y k_x k_y \right] \,,
  \label{eq:lons:m3:Hsc}
\end{align}
which are also proportional to $\mathcal{m}^\inv{m}{3}$ and an
even power of components of $\vek{k}$.  These terms are typically smaller
in magnitude than the nonrelativistic term (\ref{eq:lons:m3:Hxc}) and tend
to align the magnetic moments of the Bloch electrons perpendicular to the
local magnetic moments on the individual atoms.

A momentum-dependent spin splitting of the form $\sigma_z k_x k_y$
has recently been found in MnF$_2$ \cite{yua20}, where it was
called ``AFM-induced spin splitting''.  See also Ref.\ \cite{yua21}.
MnF$_2$ has a tetragonal rutile structure.  Its magnetic point group
is $D_{4h} (D_{2h}) = D_4 (D_2) \times \gsis$ \cite{tin64}, making
the system magnetopolar.  Similar to the hexagonal magnetic
structure discussed here, the tetragonal point group $D_{4h}
(D_{2h})$ of MnF$_2$ has $\ell_\mathrm{min}^{(\mathrm{m})} = 3$,
i.e., the lowest allowed magnetic multipole density is an
octupole~\cite{bho22a}.  While Refs.\ \cite{yua20, yua21} focused on
magnetic space groups to discuss the AFM-induced spin splitting,
magnetic point groups are sufficient to discuss this effect \cite{nye85,
tin64, bir74}.  Among the candidate materials proposed to exhibit
altermagnetism~\cite{sme22a}, CrSb has the same space group and thus
also the same point group $D_{6h} (D_{3d})$ as the magnetically
ordered structure depicted in Fig.~\ref{fig:lons}(j) and, therefore
should exhibit all of the properties discussed above.

\subsection{Magnetic quadrupolarization in lonsdaleite}
\label{sec:lons:m2}

A magnetic quadrupole density in lonsdaleite is analyzed in
Table~\ref{tab:lons:m2}.  When going from the rotation group $\Rit
\rightarrow R_i$ to $D_{6h} \times \gtis \rightarrow D_{6h}$, the
compatibility relation for a magnetic quadrupole $D_2^{--}
\rightarrow D_2^-$ reads
\begin{equation}
  R_i \mapsto D_{6h} : \quad
  D_2^- \mapsto \Gamma_1^- \oplus \Gamma_5^- \oplus \Gamma_6^- \,,
\end{equation}
so that the quadrupole remains forbidden.  The lowest-order tensor
operators transforming according to the IRs $\Gamma_1^-$,
$\Gamma_5^-$, and $\Gamma_6^-$ and consistent with the signature
$--$ of a magnetic quadrupole density are listed in
Table~\ref{tab:lons:m2}.

\begin{table*}
  \caption{\label{tab:lons:m2} Irreducible representations (IRs) and
  their lowest-order representative basis functions of a magnetic
  quadrupole and polar-toroidal vector (signatures $--$) in a
  hexagonal crystalline environment.  The IRs are labeled according
  to Koster \textit{et al.}\ \cite{kos63}.  For $C_{2v}$, however,
  the main axes $x,y,z$ have been permuted cyclically.
  Basis functions shown in square brackets do not have the required
  transformation behavior under TIS but are included for
  comparison.}
  \renewcommand{\arraystretch}{1.2}
  \let\mc\multicolumn
  \begin{tabular*}{\linewidth}{Ls{1em}EC*{4}{CC}v*{5}{C}}
    \hline \hline \rule{0pt}{2.8ex}
    \Rit \rightarrow R_i &
    \mc{9}{Cv}{D_2^{--} \rightarrow D_2^-} &
    \mc{5}{C}{D_1^{--} \rightarrow D_1^-} \\[0.5ex] \hline \rule{0pt}{2.8ex}
    D_{6h} \times \gtis \rightarrow D_{6h} & \Gamma_1^- & \oplus &
    \mc{3}{C}{\Gamma_5^-} & \oplus & \mc{3}{Cv}{\Gamma_6^-}
    & \Gamma_2^- & \oplus & \mc{3}{C}{\Gamma_5^-} \\
    & (k_x^2 - 3 k_y^2) (k_y^2 - 3 k_x^2) k_x k_y k_z & 
    & \mc{3}{C}{k_x, k_y} & &
    \mc{3}{Cv}{(k_x-ik_y)^2 k_z, (k_x+ik_y)^2 k_z} & k_z &
    & \mc{3}{C}{k_x, k_y} \\[0.5ex]
    D_{6h} (D_6) \rightarrow D_6 & \Gamma_1 & &
    \mc{3}{C}{\Gamma_5} & & \mc{3}{Cv}{\Gamma_6}
    & \Gamma_2 & & \mc{3}{C}{\Gamma_5} \\
    & (k_x^2 - 3 k_y^2) (k_y^2 - 3 k_x^2) k_x k_y k_z &
    & \mc{3}{C}{k_x, k_y} & &
    \mc{3}{Cv}{(k_x-ik_y)^2 k_z, (k_x+ik_y)^2 k_z} &
    k_z & & \mc{3}{C}{k_x, k_y} \\[0.5ex]
    D_{2h} (C_{2v}) \rightarrow C_{2v} & \Gamma_3 & &
    \Gamma_1 & \oplus & \Gamma_2 & &
    \Gamma_3 & \oplus & \Gamma_4 & \Gamma_4 & & \Gamma_1 & \oplus & \Gamma_2 \\
    & k_x k_y k_z & & k_x; [k_y^2; k_z^2] & & k_y & & k_x k_y k_z & & k_z &
    k_z & & k_x & & k_y \\
    \hline \hline
  \end{tabular*}
\end{table*}

If the four equivalent atoms in the unit cell of lonsdaleite possess
oppositely oriented local magnetic moments pointing parallel to the
lonsdaleite main axis [Fig.~\ref{fig:lons}(k)], the system acquires
a magnetic quadrupole density that yields the magnetic space group
$P6_3/m'm'c'$ (No.~194.271), and the point group symmetry is reduced
from $D_{6h} \times \gtis = D_6 \times \gsistis$ to $D_{6h} (D_6) =
D_6 \times \gstis$.  The system is thus antimagnetopolar.  More
specifically, it is the $m=0$ component of the quadrupole density that
becomes nonzero and transforms according to $\Gamma_1$ of $D_{6h}
(D_6)$.  The compatibility relations and lowest-order tensor operators for
these IRs are listed in Table~\ref{tab:lons:m2}.  Under $D_{6h} (D_6)$,
magnetic order is then signaled by a nonzero expectation value of
\begin{equation}
  \label{eq:lons:m2:Kz}
  I^\inv{m}{2}_{1 \|} \propto K^\inv{m}{2}_{1 \|} 
  = (k_x^2 - 3 k_y^2) (k_y^2 - 3 k_x^2) k_x k_y k_z \,.
\end{equation}

\begin{figure}[tb]
  \centering
  \includegraphics[width=0.50\linewidth]{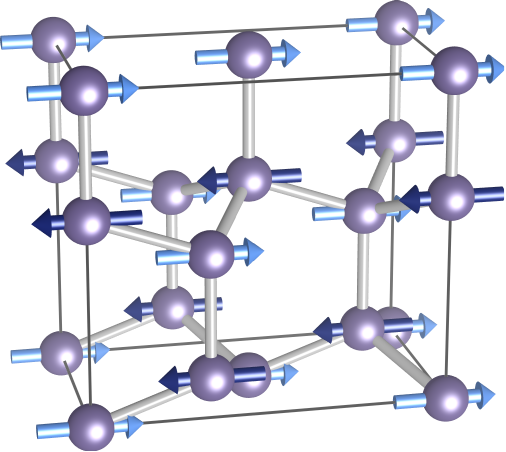}
  \caption{\label{fig:lons-x} Magnetic lonsdaleite with oppositely
  oriented local magnetic moments pointing perpendicular to the
  lonsdaleite main axis, compare Fig~\ref{fig:lons}(k).  The local
  moments give rise to a magnetic quadrupolarization ($\ell = 2$).}
\end{figure}

If, instead, the four atoms in the unit cell possess oppositely
oriented local magnetic moments pointing perpendicular to the
lonsdaleite main axis (see Fig.~\ref{fig:lons-x}), the system acquires
a magnetic quadrupole density that yields the orthorhombic space
group $Cmc'm$ (No.~63.460), and the point-group symmetry is reduced
from $D_{6h} \times \gtis$ to $D_{2h} (C_{2v})$.  The latter group
possesses only one-dimensional IRs so that the two-dimensional IRs
$\Gamma_5^-$ and $\Gamma_6^-$ of $D_{6h} \times \gtis$ split into
one-dimensional IRs of $D_{2h} (C_{2v})$.  More specifically, we
have
\begin{equation}
  D_{6h} \times \gtis \mapsto D_{2h} (C_{2v}) : \quad
  \Gamma_5^- \mapsto \Gamma_1 \oplus \Gamma_2 \,,
\end{equation}
which includes the identity representation $\Gamma_1$.  The
lowest-order tensor operators transforming according to these IRs
are listed in Table~\ref{tab:lons:m2}.  Quadrupolar magnetic order
is signaled in this case by a nonzero expectation value of
\begin{equation}
  \label{eq:lons:m2:Kx}
  I^\inv{m}{2}_{1 \perp} \propto K^\inv{m}{2}_{1 \perp} = k_x \,.
\end{equation}

Table~\ref{tab:lons:m2} also includes the compatibility relations
for a magnetotoroidal dipole ($\ell = 1$) that has the same
signature $--$ as the magnetic quadrupole density.  For the magnetic
order depicted in Fig.~\ref{fig:lons}(k) when the symmetry group is
$D_{6h} (D_6)$, a magnetotoroidal dipole transforms according to the
IRs $\Gamma_2$ and $\Gamma_5$ of $D_{6h} (D_6)$ so that it remains
forbidden.  If we instead have the magnetic order depicted in
Fig.~\ref{fig:lons-x} and the symmetry group is $D_{2h} (C_{2v})$, a
magnetotoroidal dipole transforms according to the IRs $\Gamma_1$,
$\Gamma_2$, and $\Gamma_4$.  The presence of magnetotoroidal order
is thus signaled by a nonzero expectation value of the same operator
(\ref{eq:lons:m2:Kx}) that also signals the presence of quadrupolar
magnetic order.

\subsection{Magnetization in lonsdaleite}
\label{sec:lons:m1}

A magnetic dipole density (magnetization) representing
ferromagnetism in lonsdaleite is analyzed in
Table~\ref{tab:lons:m1}.  Under the point group $D_{6h}$ of
nonmagnetic lonsdaleite, the dipole density transforms according to
the IRs $\Gamma_2^+$ and $\Gamma_5^+$.  Local magnetic moments on
the four atoms in the unit cell pointing parallel to the main axis
of lonsdaleite [Fig.~\ref{fig:lons}(l)] yield the magnetic space group
$P6_3/mm'c'$ (No.~194.270), and the point group becomes $D_{6h}
(C_{6h}) = D_6 (C_6) \times \gsis$.  The system is thus
magnetopolar.  As to be expected, the spin operator $(\hbar/2)
\sigma_z$ transforms according to $\Gamma_1^+$, and a nonzero
expectation value of $\sigma_z$ signals the presence of
ferromagnetic order.  A magnetization pointing perpendicular to the
main axis of lonsdaleite can be discussed similarly.

\subsection{Correspondence between electric and
magnetic order}
\label{sec:lons:e-m}

Figure~\ref{fig:lons} shows that, for each $\ell$, the macroscopic
multipole densities are realized by the same spatial pattern of
electric (central row) and magnetic (bottom row) atomic multipole
moments.  The close connection between electric and magnetic order
is also reflected by the crystallographic point groups characterizing
the different structures (Table~\ref{tab:symmetries}).  For odd
$\ell$, the point group characterizing the magnetopolar case is
obtained from the group characterizing the electropolar case by
replacing space inversion $i$ by time inversion $\theta$.  On the
other hand, the space group symmetries of the structures considered
in Fig.~\ref{fig:lons} are quite different; all magnetic structures
in the bottom row of Fig.~\ref{fig:lons} have nonsymmorphic space
groups, whereas pristine lonsdaleite is the only nonsymmorphic space
group in the central row.

The only exception to the correspondence between electric and
magnetic order occurs for $\ell = 4$, when antimagnetopolar order
with $\ell = 4$ can be realized as shown in Fig.~\ref{fig:lons}(i)
(all magnetic multipole densities with $\ell \le 3$ vanish for that
structure), whereas the analogous electrically ordered structure
shown in Figs.~\ref{fig:lons}(a) and~\ref{fig:lons}(e) (space group
$D_{3d}^3$, No.~164, point group $D_{3d} = D_3 \times \gsis$)
possesses not only an electric hexadecapolarization but also a
quadrupolarization as in pristine lonsdaleite
[Figs.~\ref{fig:lons}(c) and~\ref{fig:lons}(g)].  This is the reason
why panels (a) and (e) have been separated in Fig.~\ref{fig:lons}.
In fact, only cubic crystal structures do not permit an electric
quadrupolarization \cite{nye85}.

\section{Diamond family}
\label{sec:diamond}

The general theory developed in Sec.~\ref{sec:theory} is further
elucidated by applying it to electric and magnetic order in the variants of
diamond listed in Table~\ref{tab:symmetries}.  Section~\ref{sec:diamond:e4}
focuses on the electric hexadecapolarization that is compatible with the
diamond structure and therefore exists in all its variations.  The properties
of electric octupolarization, quadrupolarization and dipolarization are
discussed in Secs.~\ref{sec:diamond:e3}, \ref{sec:diamond:e2} and
\ref{sec:diamond:e1}, respectively.  As the concept of quasivectors turns out
to be useful for understanding quadrupolarizations in diamond,
Sec.~\ref{sec:quasi-vec} has been inserted to provide relevant details.
Magnetopolarizations in diamond are considered in
Secs~\ref{sec:diamond:m3} and \ref{sec:diamond:m1}, with
antimagnetopolarizations covered in Secs~\ref{sec:diamond:m4} and
\ref{sec:diamond:m2}.  The multipolarization in
Ga$_{1-x}$Mn$_x$As and related (III,Mn)\=/V compounds is discussed in Sec.~\ref{sec:diamond:gammnas}.
Connections between electric and magnetic orders
are explored in the final subsection \ref{sec:diamond:e-m}.

We follow the same procedure as in Sec.~\ref{sec:lons} for analyzing
electric order, i.e., TIS is ignored.  Again, IRs of magnetic point groups
are found by referring to their relevant nonmagnetic subgroup, and we
continue to indicate the homomorphisms relating the magnetic groups
with their nonmagnetic subgroups by `$\rightarrow$'. We also adhere to
the same general outline of each subsection as described in the preamble
of Sec.~\ref{sec:lons}.

\subsection{Electric hexadecapolarization in pristine diamond}
\label{sec:diamond:e4}

\begin{figure}[tb]
  \centering
  \includegraphics[width=0.95\linewidth]{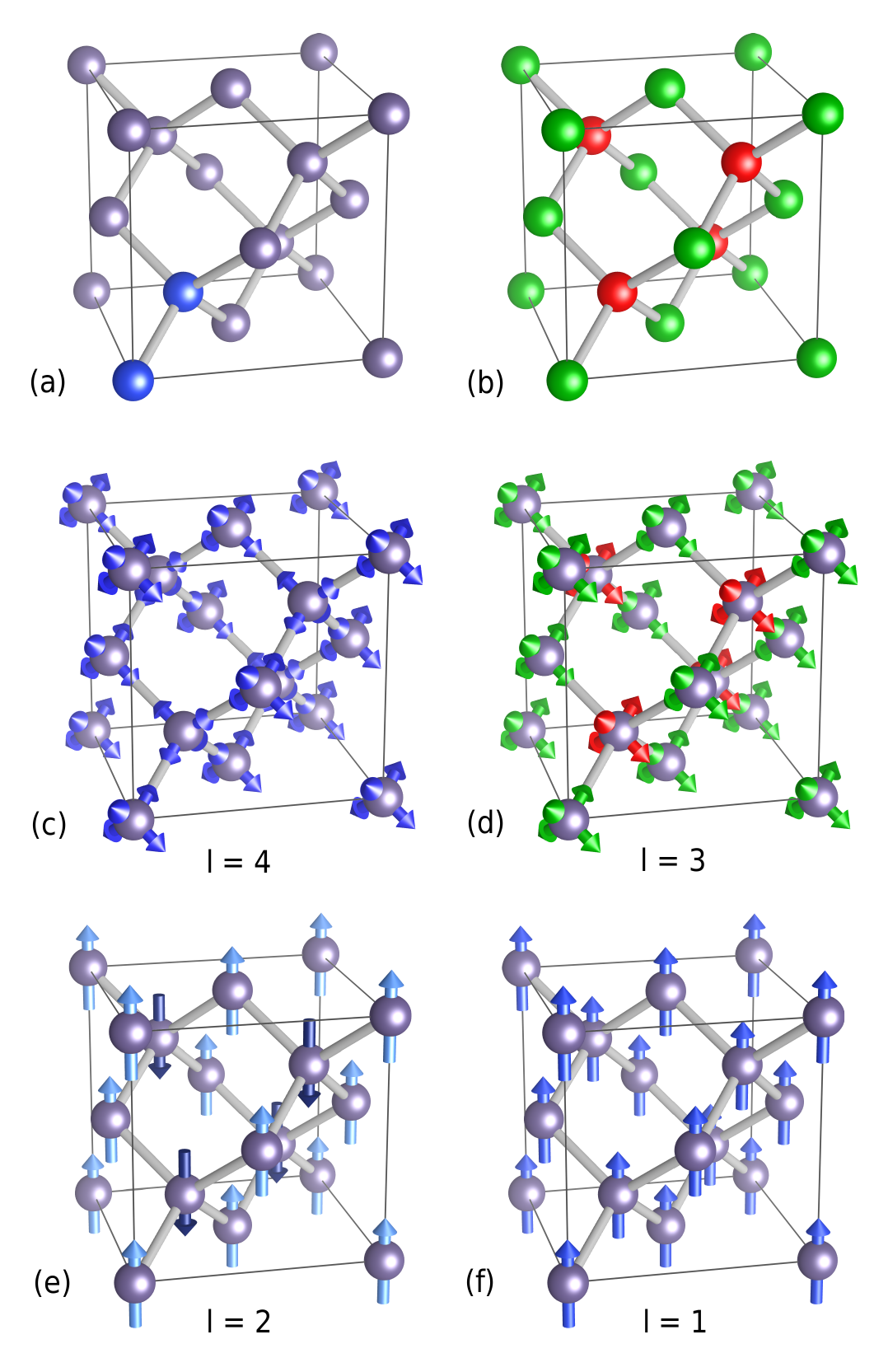}
  \caption{\label{fig:diamond-cub} Multipole densities in the
  diamond family.  Crystal structure of (a) pristine diamond and (b)
  zincblende.  In (a), the two atoms in a unit cell are highlighted
  in blue.  Local octupole moments give rise to (c) a
  hexadecapolarization ($\ell = 4$) and (d) an octupolarization
  ($\ell = 3$).  Local dipole moments give rise to (e) a
  quadrupolarization ($\ell = 2$) and (f) a dipolarization ($\ell =
  1$).}
\end{figure}

The cubic nonsymmorphic diamond structure is shown in
Fig.~\ref{fig:diamond-cub}(a).  Ignoring TIS, the space group of
diamond is $O_h^7$ (No.~227, $Fd\bar{3}m$) and the crystallographic
point group is $O_h = O \times \gsis$, i.e., diamond is parapolar.
The lowest nonvanishing electric multipole density allowed by $O_h$
is an electric hexadecapole density ($\ell = 4$) with compatibility
relation \cite{kos63}
\begin{equation}
  R_i \mapsto O_h : \quad
  D_4^+ \mapsto \Gamma_1^+ \oplus \Gamma_3^+
  \oplus \Gamma_4^+ \oplus \Gamma_5^+ \,.
\end{equation}
In lowest order of the wave vector $\vek{k}$, the scalar operator
associated with the hexadecapole density is the term
\begin{equation}
  \label{eq:diamond:e4:H}
  H^\inv{e}{4} = a^\inv{e}{4} \mathcal{m}^\inv{e}{4}
  \left( k_x^2 k_y^2 + k_y^2 k_z^2 + k_z^2 k_x^2 \right)
\end{equation}
that represents the warping of the energy dispersion $E_\sigma
(\vek{k})$ of band electrons in the cubic diamond structure
\cite{kan57, roe84}.

The site symmetry of the atoms forming the diamond structure is the
group $T_d$ \cite{hah05} that permits a local electric octupole
moment \cite{kos63}.  Similar to lonsdaleite
(Sec.~\ref{sec:lons:e2}), the octupole moment is naturally realized
by the $sp^3$ hybrid orbitals with which elements like C, Si, and Ge
form the diamond structure [Fig.~\ref{fig:diamond-cub}(c)].
Therefore, the invariant (\ref{eq:diamond:e4:H}) exists already in a
simple $sp^3$ TB model.  Using the notation of Refs.~\cite{cha75,
cha77}, in lowest order of the TB matrix elements $V_{xy}$ and
$V_{sp}$, we have $a^\inv{e}{4} \mathcal{m}^\inv{e}{4} \propto
V_{xy}^2 \, V_{sp}^2$.

The lowest electrotoroidal moment permitted in diamond
has $\ell = 9$.

\subsection{Electric octupolarization in diamond -- zincblende}
\label{sec:diamond:e3}

The unit cell of pristine diamond contains two identical atoms
[Fig.~\ref{fig:diamond-cub}(a)].  If these two atoms are distinct,
we obtain the zincblende structure [Fig.~\ref{fig:diamond-cub}(b)]
that is realized by several III\=/V semiconductors including GaAs
and InSb.  The space group of zincblende is $T_d^2$ (No.~216,
$F\bar{4}3m$), and the crystallographic point group is $T_d$.
Unlike $O_h = T_d \times \gsis$, the group $T_d$ breaks SIS so that
zincblende is electropolar.  More specifically, the compatibility
relation \cite{kos63}
\begin{equation}
  R_i \mapsto T_d : \quad
  D_3^- \mapsto \Gamma_1 \oplus \Gamma_5 \oplus \Gamma_4
\end{equation}
indicates that zincblende naturally permits an electric octupole
density ($\ell = 3$).  The corresponding
invariant in the Hamiltonian (to lowest order in~$\vek{k}$) is
\begin{equation}
  \label{eq:diamond:e3:H}
  H^\inv{e}{3} = a^\inv{e}{3} \mathcal{m}^\inv{e}{3}
  \left[ \sigma_x k_x (k_y^2 - k_z^2) + \cp \right] \,,
\end{equation}
where ``$\cp$'' denotes cyclic permutation of the preceding term.
The invariant $H^\inv{e}{3}$ represents a spin-orbit coupling
commonly known as Dresselhaus term \cite{dre55}.

\begin{table*}
  \caption{\label{tab:diamond:e3} \label{tab:diamond:m3} Irreducible
  representations (IRs) and their lowest-order representative basis
  functions of an electric (signature $-+$) and a magnetic
  (signature $+-$) octupole in a cubic crystalline environment.  The
  IRs are labeled according to Koster \textit{et al.}\ \cite{kos63}.  Basis
  functions listed for the IRs of $O_h$ are also basis functions for
  the respective IRs of $T_d$.  ``cp'' denotes the cyclic
  permutation of the preceding term.}
  \renewcommand{\arraystretch}{1.2}
  \let\mc\multicolumn
  \begin{tabular*}{\linewidth}{Ls{1em}EC*{2}{CC}}
    \hline \hline \rule{0pt}{2.8ex}
    \Rit \rightarrow R_i &
    \mc{5}{C}{D_3^{-+} \rightarrow D_3^-} \\[0.5ex] \hline \rule{0pt}{2.8ex}
    O_h \times \gtis \rightarrow O_h &
    \Gamma_2^- & \oplus & \Gamma_4^- & \oplus & \Gamma_5^- \\
    & \sigma_x k_x (k_y^2 - k_z^2) + \cp &
    & \sigma_y k_z - \sigma_z k_y, \sigma_z k_x - \sigma_x k_z,
    \sigma_x k_y - \sigma_y k_x &
    & \sigma_y k_z + \sigma_z k_y, \sigma_z k_x + \sigma_x k_z,
    \sigma_x k_y + \sigma_y k_x \\
    T_d \times \gtis \rightarrow T_d &
    \Gamma_1 & & \Gamma_5 & & \Gamma_4 \\ \hline \rule{0pt}{2.8ex}
    \Rit \rightarrow R_i &
    \mc{5}{C}{D_3^{+-} \rightarrow D_3^+} \\[0.5ex] \hline \rule{0pt}{2.8ex}
    O_h \times \gtis \rightarrow O_h &
    \Gamma_2^+ & \oplus & \Gamma_4^+ & \oplus & \Gamma_5^+ \\
    & \sigma_x k_y k_z + \cp &
    & \sigma_x, \sigma_y, \sigma_z &
    & \sigma_x (k_y^2 - k_z^2), \sigma_y (k_z^2 - k_x^2), 
    \sigma_z (k_x^2 - k_y^2); \\
    O_h (T_h) \rightarrow T_h & \Gamma_1^+ & & \Gamma_4^+ & & \Gamma_5^+
    \\ \hline \hline
  \end{tabular*}
\end{table*}

A more complete analysis of the electric octupole density in
zincblende is given in Table~\ref{tab:diamond:e3}.  Ignoring TIS,
the octupole density transforms according to $D_3^-$ of $R_i$.  A
cubic environment (point group $O_h$) yields the compatibility
relation \cite{kos63}
\begin{equation}
  R_i \mapsto O_h : \quad
  D_3^- \mapsto \Gamma_2^- \oplus \Gamma_4^- \oplus \Gamma_5^- \,,
\end{equation}
so that, as to be expected, an octupole density is forbidden in
diamond.  The lowest-order tensor operator with signature $-+$ and
transforming like $\Gamma_2^-$ is
\begin{equation}
  \label{eq:diamond:e3:K}
  K^\inv{e}{3}_{2-} = \sigma_x k_x (k_y^2 - k_z^2) + \cp \,.
\end{equation}
Examples of tensor operators transforming like $\Gamma_4^-$ and
$\Gamma_5^-$ are listed in Table~\ref{tab:diamond:e3}.
The expectation value of these operators must thus vanish in
diamond.  When the symmetry is further reduced to $T_d$
(zincblende), we get the compatibility relation \cite{kos63}
\begin{equation}
  O_h \mapsto T_d : \quad
  \Gamma_2^- \mapsto \Gamma_1, \quad
  \Gamma_4^- \mapsto \Gamma_5, \quad
  \Gamma_5^- \mapsto \Gamma_4 \,,
\end{equation}
so that the tensor operator (\ref{eq:diamond:e3:K}) becomes allowed and
yields the Dresselhaus term (\ref{eq:diamond:e3:H}).

The site symmetry of the atoms is, once again, $T_d$ \cite{hah05}.
However, similar to wurtzite, the two distinct atoms in
Fig.~\ref{fig:diamond-cub}(b) have different Wyckoff positions, and
they may carry different electric octupole moments.  Again, we decompose
these moments $\vekc{M} \equiv \vekc{M}_0 + \Delta\vekc{M}$ into one
part $\vekc{M}_0$ that is equal by symmetry for all atoms as in diamond.  These
moments thus have the same observable effect as the local moments in
pristine diamond [Fig.~\ref{fig:diamond-cub}(c)], i.e., they give rise to the
invariant (\ref{eq:diamond:e4:H}).  It is the remaining part $\Delta
\vekc{M}$ oriented oppositely that is shown in
Fig.~\ref{fig:diamond-cub}(d) and that gives rise to the Dresselhaus
term (\ref{eq:diamond:e3:H}).

The lowest electrotoroidal moment permitted in zincblende
has $\ell = 6$.

The standard $sp^3$ TB model for zincblende \cite{cha75, cha77}
provides an explicit model for Dresselhaus spin-orbit coupling
(\ref{eq:diamond:e3:H}) and its relation to the octupolarization in
zincblende.  The TB model accounts for the different atomic species
constituting zincblende structures with different on-site energies
for anions and cations; we denote their difference with $\Delta
E^\mathrm{ac}_j$, $j = s,p$.  Also, we get different overlap matrix
elements $V_{sp\sigma}$ between, on the one hand, the anion $s$ and
cation $p$ orbitals and, on the other hand, the cation $s$ and anion
$p$ orbitals.  We denote the difference between these overlap matrix
elements by $\Delta V_{sp\sigma}^\mathrm{ac}$.  This quantity
represents the scalar component (IR $\Gamma_1$) of an octupolar
charge transfer between anions and cations in zincblende.  (Larger
TB models may include multiple overlap matrix elements that permit
such an interpretation.)
In a perturbative expansion of the $sp^3$ TB model about $k=0$,
Dresselhaus spin splitting is linearly proportional to $\Delta
V_{sp\sigma}^\mathrm{ac}$, whereas it is only quadratically
proportional to $\Delta E^\mathrm{ac}_i$.
This is consistent with the $\vek{k} \cdot \vek{p}$ theory for
Dresselhaus spin-orbit coupling, where it is well-known that a
minimal model for Dresselhaus spin-orbit coupling in the lowest
conduction band (which has $s$-like symmetry) must include the
top-most valence band (consisting of $p$-bonding states in TB
language) as well as the lowest excited conduction band (consisting
of $p$-antibonding states) \cite{roe84, car88}.

The Dresselhaus term (\ref{eq:diamond:e3:H}) couples the orbital
motion of the Bloch electrons to the spin degree of freedom,
consistent with the analysis in Table~\ref{tab:tensor-op-power} that
applies to spinful models.  In spinless models of common zincblende
semiconductors, the topmost valence band originating from
$p$-bonding atomic orbitals is threefold degenerate at $k=0$.  Here,
the electric octupolarization in zincblende breaking SIS manifests
itself via terms in the perturbative expansion of the band structure
that include odd powers of the wave vector $\kk$ \cite{win03}.

\subsection{Quasivectors under point group $O_h$}
\label{sec:quasi-vec}

\begin{table*}
  \caption{\label{tab:quasi-vec} IRs of vectors and quasivectors
  under the group $O_h$ \cite{kos63}.  We list the lowest-degree
  polynomials in wave vector $\vek{k}$ and spin $\vek{\sigma}$ that
  are even and odd under TIS and that transform irreducibly
  according to these IRs.}
  \renewcommand{\arraystretch}{1.2}
  \begin{tabular*}{1.00\linewidth}{ls{0.5em}ECL}
    \hline \hline \rule{0pt}{2.8ex}%
    axial vectors & \Gamma_4^+ & \sigma_x, \sigma_y, \sigma_z; 
    k_y k_z (k_y^2-k_z^2), k_z k_x (k_z^2-k_x^2), k_x k_y (k_x^2-k_y^2) \\
    axial quasivectors & \Gamma_5^+ & k_y k_z, k_z k_x, k_x k_y;
    \sigma_x (k_y^2 - k_z^2), \sigma_y (k_z^2 - k_x^2), 
    \sigma_z (k_x^2 - k_y^2); \\ & &
    k_x (\sigma_y k_y - \sigma_z k_z), k_y (\sigma_z k_z - \sigma_x k_x),
    k_z (\sigma_x k_x - \sigma_y k_y) \\
    polar vectors & \Gamma_4^- & k_x, k_y, k_z;
    \sigma_y k_z - \sigma_z k_y, \sigma_z k_x - \sigma_x k_z,
    \sigma_x k_y - \sigma_y k_x \\
    polar quasivectors & \Gamma_5^- & 
    \sigma_y k_z + \sigma_z k_y, \sigma_z k_x + \sigma_x k_z,
    \sigma_x k_y + \sigma_y k_x;
    k_x (k_y^2-k_z^2), k_y (k_z^2-k_x^2), k_z (k_x^2-k_y^2) \\ \hline \hline
  \end{tabular*}
\end{table*}  

Before discussing electric quadrupole densities in diamond, we
introduce the concept of \emph{quasivectors} in systems with point
group $O_h$.  The point group $O_h$ has four three-dimensional IRs
denoted $\Gamma_4^\pm$ and $\Gamma_5^\pm$ in Koster's notation
\cite{kos63}.  Function triples like the components $k_x, k_y, k_z$
of the wave vector $\vek{k}$ that behave like a polar vector under
all symmetry elements of $O_h$ are ascribed the IR $\Gamma_4^-$ of
$O_h$.  On the other hand, function triples such as
$k_x(k_y^2-k_z^2), k_y(k_z^2-k_x^2), k_z(k_x^2-k_y^2)$ transform
according to the IR $\Gamma_5^-$ of $O_h$.  The latter functions
behave like polar vectors under half of the symmetry elements of
$O_h$.  However, they change sign under proper $\pm\pi/2$ rotations
about axes $\avg{100}$ and proper $\pi$ rotations about axes
$\avg{110}$, while they behave like axial vectors (not changing sign)
under improper $\pm\pi/2$ rotations about axes $\avg{100}$ and
improper $\pi$ rotations about axes $\avg{110}$.  We call sets of
functions transforming according to $\Gamma_5^-$ of $O_h$
\emph{polar quasivectors}.

Similarly, function triples like $k_y k_z (k_y^2-k_z^2), k_z
k_x(k_z^2-k_x^2),k_x k_y (k_x^2-k_y^2)$ behaving like axial vectors
under all symmetry elements of $O_h$ are ascribed the IR
$\Gamma_4^+$ of $O_h$.  On the other hand, function triples such as
$k_y k_z, k_z k_x, k_x k_y$ transforming according to the IR
$\Gamma_5^+$ of $O_h$ behave like axial vectors under half of the
symmetry elements of $O_h$.  However, they change sign under proper
$\pm\pi/2$ rotations about axes $\avg{100}$ and proper $\pi$
rotations about axes $\avg{110}$, while they behave like polar
vectors (changing sign) under improper $\pm\pi/2$ rotations about
axes $\avg{100}$ and improper $\pi$ rotations about axes $\avg{110}$.
We call sets of functions transforming according to $\Gamma_5^+$ of
$O_h$ \emph{axial quasivectors}.

The IRs for vectors and quasivectors under the point group $O_h$
are summarized in Table~\ref{tab:quasi-vec}.  We illustrate these
IRs with representative basis functions transforming according to
these IRs.  The table gives the lowest-degree polynomials in wave
vector $\vek{k}$ and spin $\vek{\sigma}$ that are even and odd under
TIS.  By definition, in a cubic environment (point group $O_h$) none
of these vectors are observable.  They become observable when the
crystal symmetry is reduced.

\subsection{Electric quadrupolarization in diamond}
\label{sec:diamond:e2}

Next we discuss an electric quadrupole density in diamond.  The
analysis is summarized in Table~\ref{tab:diamond:e2}.  The cubic
group $O_h$ yields the compatibility relation \cite{kos63}
\begin{equation}
  \label{eq:diamond:e2:R-O}
  R_i \mapsto O_h : \quad
  D_2^+ \mapsto \Gamma_3^+ \oplus \Gamma_5^+ \,,
\end{equation}
i.e., the five components of the quadrupole decompose into an axial
quasivector ($\Gamma_5^+$) and two components transforming
according to $\Gamma_3^+$.  As to be expected, an electric
quadrupole is forbidden for $O_h$.

\begin{table*}
  \caption{\label{tab:diamond:e2} Irreducible representations (IRs)
  and their lowest-order representative basis functions of an
  electric quadrupole and axial-toroidal vector (signatures $++$) in
  a cubic crystalline environment.  The IRs are labeled according to
  Koster \textit{et al.}\ \cite{kos63}.}
  \renewcommand{\arraystretch}{1.2}
  \let\mc\multicolumn
  \begin{tabular*}{\linewidth}{Ls{-1.0em}EC*{3}{CC}v*{3}{C}}
    \hline \hline  \rule{0pt}{2.8ex}
    \Rit \rightarrow R_i &
    \mc{7}{Cv}{D_2^{++} \rightarrow D_2^+} &
    \mc{3}{C}{D_1^{++} \rightarrow D_1^+} \\[0.5ex] \hline \rule{0pt}{2.8ex}
    O_h \times \gtis \rightarrow O_h & \mc{3}{C}{\Gamma_3^+} & \oplus &
    \mc{3}{Cv}{\Gamma_5^+} & \mc{3}{C}{\Gamma_4^+} \\
    & \mc{3}{C}{(2k_z^2-k_x^2-k_y^2), \sqrt{3} (k_x^2-k_y^2)} & &
    \mc{3}{Cv}{k_y k_z, k_z k_x, k_x k_y} &
    \mc{3}{C}{k_y k_z (k_y^2-k_z^2), k_z k_x(k_z^2-k_x^2),
    k_x k_y (k_x^2-k_y^2)} \\[0.5ex]
    {}[001]: D_{4h} & \Gamma_1^+ & \oplus & \Gamma_3^+ & &
    \Gamma_4^+ & \oplus & \Gamma_5^+ &
    \Gamma_2^+ & \oplus & \Gamma_5^+ \\
    & k_x^2+k_y^2; k_z^2 & & k_x^2-k_y^2 & & k_x k_y & &
    k_y k_z, k_z k_x & k_x k_y (k_x^2-k_y^2) & &
    k_y k_z (k_y^2-k_z^2), k_z k_x(k_z^2-k_x^2) \\[0.5ex]
    {}[mmn]: C_{2h} & \Gamma_1^+ & \oplus & \Gamma_2^+ & &
    2\,\Gamma_1^+ & \oplus & \Gamma_2^+ &
    \Gamma_1^+ & \oplus & 2 \, \Gamma_2^+ \\
    & k_x^2; k_y^2; k_z^2; k_z k_x & & k_x k_y; k_y k_z & &
    k_x^2; k_y^2; k_z^2; k_z k_x & & k_x k_y; k_y k_z &
    k_x^2; k_y^2; k_z^2; k_z k_x & & k_x k_y; k_y k_z \\
    \hline \hline
 \end{tabular*}
\end{table*}

As in pristine lonsdaleite (Sec.~\ref{sec:lons:e2}), we consider the
case that the nonzero component of the quadrupole density is $m=0$.
If the main axis of the quadrupole is parallel to the
crystallographic $[001]$ axis, the point group becomes $D_{4h}$ and
we get the compatibility relations
\begin{equation}
  O_h \mapsto D_{4h} : \left\{
  \begin{array}{rcls{0.6em}rcl}
    \Gamma_3^+ & \mapsto & \Gamma_1^+ \oplus \Gamma_3^+ \\[0.5ex]
    \Gamma_5^+ & \mapsto & \Gamma_4^+ \oplus \Gamma_5^+ \,.
  \end{array}
  \right.
\end{equation}
In this case we obtain the same invariant as in lonsdaleite,
\begin{subequations}
  \begin{align}
    H^\inv{e}{2}
    & = a^\inv{e}{2} \mathcal{m}^\inv{e}{2}_z K^\inv{e}{2}_1 \\
    & = a^\inv{e}{2} \mathcal{m}^\inv{e}{2}_z (k_x^2 + k_y^2 - 2 \, k_z^2) \,,
  \end{align}
\end{subequations}
that represents an effective-mass anisotropy in the energy
dispersion $E_\sigma (\vek{k})$ of band electrons in diamond.  The
axial quasivector transforming according to $\Gamma_5^+$ of $O_h$
remains forbidden.

We compare with the case that the main axis of the quadrupole is
parallel to the crystallographic $[mmn]$ axis, when the point group
symmetry becomes $C_{2h}$ and we get the compatibility relations
\begin{equation}
  \label{eq:diamond:e2:O-C2h}
  O_h \mapsto C_{2h} : \left\{
  \begin{array}{rcls{0.6em}rcl}
    \Gamma_3^+ & \mapsto & \Gamma_1^+ \oplus \Gamma_2^+ \\[0.5ex]
    \Gamma_5^+ & \mapsto & 2 \Gamma_1^+ \oplus \Gamma_2^+ \,.
  \end{array}
  \right.
\end{equation}
Here the IR $\Gamma_1$ appears three times so that the
quadrupole density has three independent components, two of which
represent the axial quasivector ($\Gamma_5^+$) in the decomposition
(\ref{eq:diamond:e2:R-O}) of the quadrupole density.

Table~\ref{tab:diamond:e2} also includes the compatibility relations
for an electrotoroidal dipole ($\ell = 1$) that has the same
signature $++$ as the electric quadrupole density.  In the language
of Sec.~\ref{sec:quasi-vec}, the electrotoroidal dipole is an axial
vector, whereas the electric quadrupole density includes a part
transforming like an axial quasivector [Eq.\
(\ref{eq:diamond:e2:R-O})].  When the symmetry is reduced from $O_h$
to $D_{4h}$, the electrotoroidal dipole remains forbidden.  On the
other hand, the reduced symmetry $C_{2h}$ implies that not only the
axial quasivector ($\Gamma_5^+$ of $O_h$) becomes allowed for
$C_{2h}$ [Eq.\ (\ref{eq:diamond:e2:O-C2h})], but also an axial
vector ($\Gamma_4^+$ of $O_h$) becomes
observable
\begin{equation}
  O_h \mapsto C_{2h} : \quad
  \Gamma_4^+ \mapsto \Gamma_1^+ \oplus 2 \, \Gamma_2^+ \,.
\end{equation}
It becomes clear from Table~\ref{tab:diamond:e2} that the reason for
the observability of both quantities under $C_{2h}$ lies in the fact
that axial vectors and axial quasivectors are only distinct
quantities under the high symmetry of the point group $O_h$.  But
they represent the same observable physics when the symmetry is
reduced to a group like $C_{2h}$ that makes both of these quantities
measurable.  Both quantities manifest themselves via terms in the
energy dispersion of band electrons proportional to the invariants
$k_z k_x$ and $k_z k_x(k_z^2-k_x^2)$.

\subsubsection{Strain}

An electric quadrupole density becomes allowed in diamond when the
symmetry is reduced by means of strain \cite{suz74} or quantum
confinement \cite{win03}.  Similar to an electric quadrupole
density, in cartesian coordinates strain is characterized via a
symmetric rank-2 tensor $\epsilon$ \cite{lan7e}.  The trace of
$\epsilon$ represents the effect of hydrostatic pressure.  Ignoring
hydrostatic pressure, the tensor $\epsilon$ is traceless.  The
effect of strain \cite{bir74} is then equivalent to inducing an
electric quadrupole density.  A nonzero component $\epsilon_{ij}$
implies that the respective component of the electric quadrupole
density becomes observable, too.

\begin{figure}[tb]
  \centering
  \includegraphics[width=0.99\linewidth]{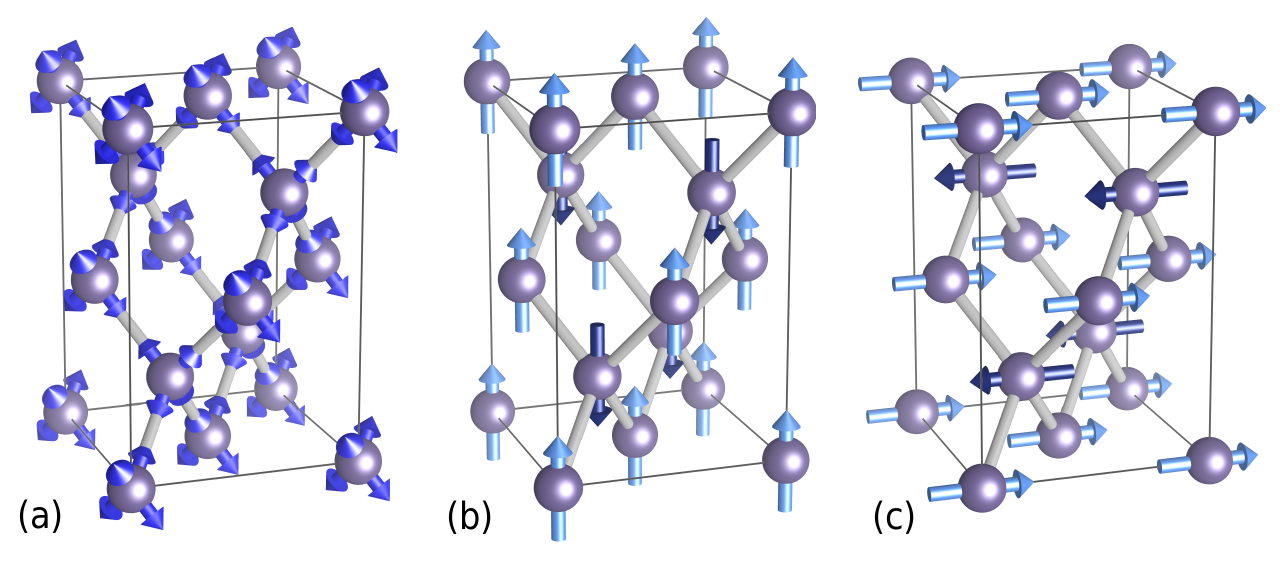}
  \caption{\label{fig:diamond-tetra} Quadrupole densities in
  tetragonally distorted diamond.  (a) The distorted electric
  octupole moments due to the $sp^3$ hybrid orbitals give rise to
  local electric quadrupole moments which, in turn, give rise to an
  electric quadrupole density.  Alternating patterns of magnetic
  dipole moments oriented (b) parallel and (c) perpendicular to the
  tetragonal axis give rise to magnetic quadrupole densities.}
\end{figure}

The strain due to uniaxial stress applied in $[001]$ direction
reduces the space group of diamond to the tetragonal group
$D_{4h}^{19}$ (No.~141, $I4_1/amd$) with point group $D_{4h}$.  The
site symmetry of the atoms becomes $D_{2d}$ that supports an
electric quadrupole moment [Fig.~\ref{fig:diamond-tetra}(a)].

\subsection{Electric dipolarization in zincblende}
\label{sec:diamond:e1}

Similar to wurtzite, an electric dipole density becomes allowed if
the symmetry of diamond is reduced from $O_h$ to one of the polar
subgroups of $O_h$ including $C_{4v}$, $C_{3v}$, and $C_{2v}$.  This
is summarized in Table~\ref{tab:diamond:e1}.  In these cases, the
dipole density manifests itself via the same Rashba term
(\ref{eq:lons:e1:H}) as in wurtzite.  Starting from a bulk
zincblende structure, the polar point groups $C_{3v}$ and $C_{2v}$
can be obtained experimentally by applying uniaxial strain in the
crystallographic direction $[111]$ ($C_{3v}$) or $[110]$ ($C_{2v}$),
which is the familiar piezoelectric effect \cite{nye85} that exists
for zincblende structures (but not for diamond).  It has been noted
previously \cite{dya86, car05} that for systems with point groups
$C_{3v}$ and $C_{2v}$, the Dresselhaus term takes the form of a
Rashba term linear in the wave vector $\vek{k}$.  Indeed, this is
due to the fact that the system becomes polar and possesses an
electric dipole density.

\begin{table}
  \caption{\label{tab:diamond:e1} \label{tab:diamond:m1} Irreducible
  representations (IRs) and their lowest-order representative basis
  functions of an electric dipole (signature $-+$) and magnetic
  dipole (signature $+-$) in a cubic crystalline environment.  The
  IRs are labeled according to Koster \textit{et al.}\
  \cite{kos63}.}
  \renewcommand{\arraystretch}{1.2}
  \let\mc\multicolumn
  \tabcolsep 0.15em
  \begin{tabular*}{\linewidth}{Ls{0.5em}ECCCCC}
    \hline \hline  \rule{0pt}{2.8ex}
    \Rit \rightarrow R_i &
    \mc{5}{C}{D_1^{-+} \rightarrow D_1^-} \\[0.5ex] \hline \rule{0pt}{2.8ex}
    O_h \times \gtis \rightarrow O_h & \mc{5}{C}{\Gamma_4^-} \\
    & \mc{5}{C}{\sigma_y k_z - \sigma_z k_y,
    \sigma_z k_x - \sigma_x k_z, \sigma_x k_y - \sigma_y k_x} \\[0.5ex]
    T_d \times \gtis \rightarrow T_d & \mc{5}{C}{\Gamma_5} \\
    & \mc{5}{C}{\sigma_y k_z - \sigma_z k_y,
    \sigma_z k_x - \sigma_x k_z, \sigma_x k_y - \sigma_y k_x} \\[0.5ex]
    {} [001] : C_{4v} % \times \gtis \rightarrow C_{4v}
    & \Gamma_1 & \oplus & \mc{3}{C}{\Gamma_5} \\
    & \sigma_x k_y - \sigma_y k_x & &
    \mc{3}{C}{\sigma_y k_z - \sigma_z k_y, \sigma_z k_x - \sigma_x k_z} \\
    {} [111] : C_{3v} % \times \gtis \rightarrow C_{3v} 
    & \Gamma_1 & \oplus & \mc{3}{C}{\Gamma_3} \\
    & \sigma_x k_y - \sigma_y k_x & &
    \mc{3}{C}{\sigma_y k_z - \sigma_z k_y, \sigma_z k_x - \sigma_x k_z} \\
    {} [110] : C_{2v} % \times \gtis \rightarrow C_{2v}
    & \Gamma_1 & \oplus & \Gamma_2 & \oplus & \Gamma_4 \\
    & \sigma_x k_y; \sigma_y k_x & & \sigma_y k_z; \sigma_z k_y & &
    \sigma_x k_z; \sigma_z k_x \\
  \end{tabular*} \\[0.08ex]
  \begin{tabular*}{\linewidth}{Ls{1.5em}ECCCs{3em}}
    \hline \rule{0pt}{2.8ex}
    \Rit \rightarrow R_i & &
    \makebox[0pt]{$D_1^{+-} \rightarrow D_1^+$} & \\[0.5ex]
    \hline \rule{0pt}{2.8ex}
    O_h \times \gtis \rightarrow O_h & &
    \makebox[0pt]{$\Gamma_4^+$} & \\
    & & \makebox[0pt]{$\sigma_x, \sigma_y, \sigma_z$} &  \\[0.5ex]
    {} [001] : D_{4h} (C_{4h}) \rightarrow C_{4h} &
    \Gamma_1^+ & \oplus & \Gamma_3^+ \oplus \Gamma_4^+ \\
    & \sigma_z & & \sigma_x, \sigma_y \\
    \hline \hline
  \end{tabular*}
\end{table}

\subsection{Magnetic hexadecapolarization in diamond}
\label{sec:diamond:m4}

In analogy with the electric hexadecapole density characterizing
pristine diamond (Sec.~\ref{sec:diamond:e4}) and the octupole
density characterizing the zincblende structure
(Sec.~\ref{sec:diamond:e3}), we can also discuss magnetic multipole
densities that can be modeled using atomic octupoles on the two
sublattices of diamond as the elementary building blocks for the
multipolar order \cite{kur08}.
When going from the rotation group $\Rit \rightarrow R_i$ to $O_h
\times \gtis \rightarrow O_h$, the compatibility relation for a
magnetic hexadecapole $D_4^{--} \rightarrow D_4^-$ reads
\begin{equation}
  R_i \mapsto O_h : \quad
  D_4^- \mapsto \Gamma_1^- \oplus \Gamma_3^-
  \oplus \Gamma_4^- \oplus \Gamma_5^- \,,
\end{equation}
hence the hexadecapole remains forbidden.  If local magnetic
octupoles on the two sublattices of diamond are oriented as in
Fig.~\ref{fig:diamond-cub}(c), they reduce the symmetry of diamond
to the space group $Fd'\bar{3}'m'$ (No.~227.132), and the point group
becomes
\begin{equation}
  \label{eq:diamond:m4:G-red}
  O_h (O) = O \times \gstis \rightarrow O \,,
\end{equation}
i.e., the system becomes antimagnetopolar.  More
specifically, we get the compatibility relations
\begin{equation}
  O_h \mapsto O : \quad
  \Gamma_1^- \mapsto \Gamma_1 , \;
  \Gamma_3^- \mapsto \Gamma_3 , \;
  \Gamma_4^- \mapsto \Gamma_4 , \;
  \Gamma_5^- \mapsto \Gamma_5 \,.
\end{equation}
Under $O_h (O)$, hexadecapolar magnetic order is signaled by a
nonzero expectation value of the indicator
\begin{equation}
  \label{eq:diamond:m4:K}
  I^\inv{m}{4}_1 \propto K^\inv{m}{4}_1
  = k_x k_y k_z (k_y^2-k_z^2) (k_z^2-k_x^2) (k_x^2-k_y^2) \,.
\end{equation}
Like the Dresselhaus term (\ref{eq:diamond:e3:H}),
$I^\inv{m}{4}_1$ exhibits cubic symmetry.

Incidentally, the group $O_h (O)$ also permits a magnetic monopole
density ($\ell = 0$)
\begin{equation}
  R_i \mapsto O_h \mapsto O : \quad
  D_0^- \mapsto \Gamma_1^- \mapsto \Gamma_1
\end{equation}
that gives rise to the same observable physics as the hexadecapole
density.  In Ref.\ \cite{wat18}, the indicator
(\ref{eq:diamond:m4:K}) was associated with a magnetic
monopolarization.

The site symmetry of the magnetic sites in
Fig.~\ref{fig:diamond-cub}(c) is $T_d (T)$ that supports a magnetic
octupole moment.  Similar to the invariant (\ref{eq:lons:e2:H}) in
lonsdaleite, Eq.\ (\ref{eq:diamond:m4:K}) can already be evaluated
in a simple $sp^3$ TB model for diamond \cite{cha75, cha77}.  Local
magnetic octupole moments can be implemented as a spin Zeeman term
for the four $sp^3$ orbitals on each site, assuming that the
exchange field complementing each $sp^3$ orbital is oriented along
the respective orbital.  A perturbative expansion of the resulting
TB model includes a term $\propto K^\inv{m}{4}_1$ from Eq.\
(\ref{eq:diamond:m4:K}).

A magnetic hexadecapolarization has previously been discussed for a
particular materials system \cite{wat17}.

\subsection{Magnetic octupolarization in diamond}
\label{sec:diamond:m3}

If the orientation of the atomic magnetic octupoles on one of the
diamond sublattices is reversed as in Fig.~\ref{fig:diamond-cub}(d),
these octupoles give rise to a macroscopic magnetic octupole
density whose symmetry properties are summarized in
Table~\ref{tab:diamond:m3}.
When going from the rotation group $\Rit \rightarrow R_i$ to $O_h
\times \gtis \rightarrow O_h$, the compatibility relation for a
magnetic octupole $D_3^{+-} \rightarrow D_3^+$ reads
\begin{equation}
  R_i \mapsto O_h : \quad
  D_3^+ \mapsto \Gamma_2^+ \oplus \Gamma_4^+ \oplus \Gamma_5^+ \,,
\end{equation}
hence the octupole remains forbidden.  The lowest-order tensor
operators transforming according to the IRs $\Gamma_2^+$,
$\Gamma_4^+$, and $\Gamma_5^+$ of $O_h$ and consistent with the
signature $+-$ of a magnetic octupole density are listed in
Table~\ref{tab:diamond:m3}.

The magnetic octupole density illustrated in
Fig.~\ref{fig:diamond-cub}(d) reduces the symmetry of diamond to the
space group $Fd\bar{3}m'$ (No.~227.131), and the point group becomes
\begin{equation}
  \label{eq:diamond:m3:G-red}
 O_h (T_h) = O(T) \times \gsis \rightarrow T_h = T \times \gsis  \,,
\end{equation}
i.e., the system becomes magnetopolar.  More
specifically, we get the compatibility relations
\begin{equation}
  O_h \mapsto T_h : \quad
  \Gamma_2^+ \mapsto \Gamma_1^+ , \;
  \Gamma_4^+ \mapsto \Gamma_4^+ , \;
  \Gamma_5^+ \mapsto \Gamma_5^+ \,.
\end{equation}
Therefore, under $O_h (T_h)$ octupolar magnetic order is signaled by
a nonzero expectation value of
\begin{equation}
  \label{eq:diamond:m3:K}
  I^\inv{m}{3}_1 \propto K^\inv{m}{3}_1
  = \sigma_x k_y k_z + \sigma_y k_z k_x + \sigma_z k_x k_y \,.
\end{equation}
Unlike Eq.\ (\ref{eq:lons:m3:Hxc}) in lonsdaleite, but similar to the
Dresselhaus term (\ref{eq:diamond:e3:H}), this term preserves the
cubic symmetry and can thus be viewed as a generalized form of
altermagnetism \cite{sme22a} without a global spin-quantization axis.

The site symmetry of the local magnetic moments in
Fig.~\ref{fig:diamond-cub}(d) is again $T_d (T)$.  The magnetic
octupolarization can be implemented in an $sp^3$ TB model as
described in Sec.~\ref{sec:diamond:m4} for the magnetic
hexadecapolarization, but by giving opposite signs to the local
octupole moments on the two sublattices.

\subsection{Magnetic quadrupolarization in diamond}
\label{sec:diamond:m2}

\begin{table*}
  \caption{\label{tab:diamond:m2} Irreducible representations (IRs) and their
  lowest-order representative basis functions of a magnetic quadrupole and
  polar-toroidal vector (signatures $--$) in a cubic crystalline
  environment.
  The IRs are labeled according to Koster \textit{et al.}\
  \cite{kos63}.  For $D_{2d}$, however, Koster's coordinate system
  has been rotated by $\pi/4$ about the main axis of $D_{2d}$,
  which changes the representative basis functions for the IRs
  $\Gamma_2$, $\Gamma_3$, and $\Gamma_4$ of $D_{2d}$.  For $C_{2v}$,
  the main axes $x,y,z$ have been permuted cyclically.
  Basis functions shown in square brackets do not have the required
  transformation behavior under TIS, but are included for comparison.}
  \renewcommand{\arraystretch}{1.2}
  \let\mc\multicolumn
  \begin{tabular*}{\linewidth}{Ls{1em}EC*{4}{CC}v*{5}{C}}
    \hline \hline \rule{0pt}{2.8ex}
    \Rit \rightarrow R_i &
    \mc{9}{Cv}{D_2^{--} \rightarrow D_2^-} &
    \mc{5}{C}{D_1^{--} \rightarrow D_1^-} \\[0.5ex] \hline \rule{0pt}{2.8ex}
    O_h \times \gtis \rightarrow O_h & \mc{3}{C}{\Gamma_3^-} & \oplus &
    \mc{5}{Cv}{\Gamma_5^-} & \mc{5}{C}{\Gamma_4^-} \\
    & \mc{3}{C}{(2k_z^2-k_x^2-k_y^2)k_x k_y k_z,} & &
    \mc{5}{Cv}{k_x(k_y^2-k_z^2), k_y(k_z^2-k_x^2), k_z(k_x^2-k_y^2)} & \mc{5}{C}{k_x,k_y,k_z} \\
    & \mc{3}{C}{\sqrt{3} (k_x^2-k_y^2)k_x k_y k_z} & &  & & & & & & & & & \\[0.5ex]
    D_{4h} (D_{2d}) \rightarrow D_{2d} & \Gamma_2 & \oplus & \Gamma_3 & &
    \Gamma_1 & \oplus & \mc{3}{Cv}{\Gamma_5} &
    \Gamma_4 & \oplus & \mc{3}{C}{\Gamma_5} \\
    & k_x k_y k_z & & (k_x^2-k_y^2) k_x k_y k_z & & k_z (k_x^2-k_y^2); & &
    \mc{3}{Cv}{k_x(k_y^2-k_z^2), k_y(k_z^2-k_x^2);} & k_z; & & \mc{3}{C}{k_x,k_y} \\
    & & & & &  [k_z^2; k_x^2+k_y^2] & & \mc{3}{Cv}{k_x,k_y} &
    [k_x^2-k_y^2] & & \\[0.5ex] \hline \rule{0pt}{2.8ex}
    D_{4h} \times \gtis \rightarrow D_{4h} &
    \Gamma_1^- & \oplus & \Gamma_3^- & &
    \Gamma_4^- & \oplus & \mc{3}{Cv}{\Gamma_5^-} &
    \Gamma_2^- & \oplus & \mc{3}{C}{\Gamma_5^-} \\
    & (k_x^2-k_y^2)k_x k_y k_z & & k_x k_y k_z & & k_z (k_x^2-k_y^2) & &
    \mc{3}{Cv}{k_x(k_y^2-k_z^2), k_y(k_z^2-k_x^2);} &
    k_z & & \mc{3}{C}{k_x,k_y} \\
    & & & & & & & \mc{3}{Cv}{k_x,k_y} & \\[0.5ex]
    D_{2h} (C_{2v}) \rightarrow C_{2v} & \Gamma_3 & & \Gamma_3 & & \Gamma_4 & &
    \Gamma_1 & \oplus & \Gamma_2 & \Gamma_4 & & \Gamma_1 & \oplus & \Gamma_2 \\
    & k_x k_y k_z & & k_x k_y k_z & & k_z & & k_x; [k_y^2; k_z^2] & & k_y &
    k_z & & k_x & & k_y \\
    \hline \hline
 \end{tabular*}
\end{table*}

A magnetic quadrupole density is illustrated in
Fig.~\ref{fig:diamond-cub}(e) \cite{misc:diamond-afm}.  Recently,
such a quadrupole density has been analyzed in Ref.\ \cite{win20}.
Here we present a more detailed discussion that is summarized in
Table~\ref{tab:diamond:m2}.
When going from the rotation group $\Rit \rightarrow R_i$ to $O_h
\times \gtis \rightarrow O_h$, the compatibility relation for a
magnetic quadrupole $D_2^{--} \rightarrow D_2^-$ reads
\begin{equation}
  \label{eq:diamond:m2:R-Oh}
  R_i \mapsto O_h : \quad
  D_2^- \mapsto \Gamma_3^- \oplus \Gamma_5^- \,,
\end{equation}
i.e., the quadrupole remains forbidden.  The lowest-order tensor
operators transforming according to the IRs $\Gamma_3^-$ and
$\Gamma_5^-$ and consistent with the signature $--$ of a magnetic
quadrupole density are listed in Table~\ref{tab:diamond:m2}.  In
Ref.\ \cite{win20}, the indicator associated with $\Gamma_5^-$
\begin{equation}
  \vek{I}^\inv{m}{2}_{5-} \propto \vek{K}^\inv{m}{2}_{5-} =
  \begin{pmatrix}
    k_x(k_y^2-k_z^2) \\[0.5ex] k_y(k_z^2-k_x^2) \\[0.5ex] k_z(k_x^2-k_y^2)
  \end{pmatrix}
\end{equation}
was called N\'eel operator, and it was argued that a nonzero
expectation value of this quantity signals the presence of AFM order
in the magnetic diamond structure shown in
Fig.~\ref{fig:diamond-cub}(e).  The concept of indicators introduced
in Eq.\ (\ref{eq:indicator}) extends this idea to different types of
electric and magnetic multipolar order in crystal structures.

The diamond structure with locally alternating magnetic dipoles
[Fig.~\ref{fig:diamond-cub}(e)] reduces the space group
symmetry to $I4_1'/a'm'd$ (No.~141.566), and the magnetic point
group becomes $D_{4h} (D_{2d}) = D_4 (D_2) \times \gstis \rightarrow
D_{2d}$ so that the system is antimagnetopolar.  More
specifically, we get the compatibility relations
\begin{equation}
  O_h \mapsto D_{2d} : \left\{
  \begin{array}{rcls{0.6em}rcl}
    \Gamma_3^- & \mapsto & \Gamma_3 \oplus \Gamma_3 \\[0.5ex]
    \Gamma_5^- & \mapsto & \Gamma_1 \oplus \Gamma_5  \,.
  \end{array}
  \right.
\end{equation}
Therefore, under $D_{4h} (D_{2d})$, quadrupolar magnetic order is
signaled by a nonzero expectation value of
\begin{equation}
  \label{eq:diamond:m2:K-D2d}
  D_{4h} (D_{2d}): \quad
  I^\inv{m}{2}_1 \propto K^\inv{m}{2}_1
  = k_z(k_x^2-k_y^2) \,.
\end{equation}
As discussed in Ref.~\cite{win20}, the locally alternating magnetic
dipoles [Fig.~\ref{fig:diamond-cub}(e)] can be implemented in an
$sp^3$ TB model \cite{cha75, cha77} as a local exchange field that
is oriented oppositely on the two sublattices of diamond.  A perturbative
expansion of the TB model to lowest order in $\vek{k}$ then yields a term
proportional to the right-hand-side of Eq.\ (\ref{eq:diamond:m2:K-D2d}).

\subsubsection{Magnetic quadrupolarization in quantum-confined diamond}
\label{sec:diamond:m2:qw}

As discussed in Sec.~\ref{sec:diamond:e2}, an electric quadrupole
density that may be realized via uniaxial strain or quantum
confinement can reduce the symmetry of diamond from $O_h$ to
$D_{4h}$.  Reference \cite{win20} studied a magnetic quadrupole
density for this scenario, and we want to review the case using the
language developed in the present work.  We assume that the uniaxial
strain is oriented in the crystallographic $[001]$ direction, which
is symmetry-wise equivalent to a quantum well grown on a $(001)$
surface.  
When going from the rotation group $\Rit \rightarrow R_i$ to $D_{4h}
\times \gtis \rightarrow D_{4h}$, the compatibility relation for a
magnetic quadrupole $D_2^{--} \rightarrow D_2^-$ reads
\begin{equation}
  R_i \mapsto D_{4h} : \quad
  D_2^- \mapsto \Gamma_1^- \oplus \Gamma_3^-
  \oplus \Gamma_4^- \oplus \Gamma_5^- \,.
\end{equation}
If the alternating magnetic moments on the two diamond sublattices
point parallel to the $[001]$ direction
[Fig.~\ref{fig:diamond-tetra}(b)], the magnetic point group is
reduced from $D_{4h} \times \gtis$ to once again $D_{4h} (D_{2d})$
that was already discussed above.  If the moments point parallel to
the $[100]$ direction [Fig.~\ref{fig:diamond-tetra}(c)], the
symmetry group becomes $D_{2h} (C_{2v}) \rightarrow C_{2v}$.  The
compatibility relations are
\begin{equation}
  \label{eq:diamond:m2:D4h-C2v}
  D_{4h} \mapsto C_{2v} : \left\{
  \begin{array}{rcls{0.6em}rcl}
    \Gamma_1^- & \mapsto & \Gamma_3 \\[0.5ex]
    \Gamma_3^- & \mapsto & \Gamma_3 \\[0.5ex]
    \Gamma_4^- & \mapsto & \Gamma_4 \\[0.5ex]
    \Gamma_5^- & \mapsto & \Gamma_1 \oplus \Gamma_2 \,,
  \end{array}
  \right.
\end{equation}
with representative basis functions listed in
Table~\ref{tab:diamond:m2}.  Thus under $D_{2h} (C_{2v})$,
quadrupolar magnetic order becomes once again allowed, and its
presence is signaled by a nonzero expectation value of \cite{win20}
\begin{equation}
  \label{eq:diamond:m2:K-C2v}
  D_{2h} (C_{2v}) : \quad
  I^\inv{m}{2}_1 \propto K^\inv{m}{2}_1 = k_x \,.
\end{equation}

Table~\ref{tab:diamond:m2} also includes the compatibility relations
for a magnetotoroidal dipole ($\ell = 1$) that has the same signature
$--$ as the magnetic quadrupole density.  In the language of
Sec.~\ref{sec:quasi-vec}, under $O_h \times \gtis \rightarrow O_h$
the magnetotoroidal dipole is a polar vector (IR $\Gamma_4^-$),
whereas the magnetic quadrupole density includes a part transforming
like a polar quasivector [IR $\Gamma_5^-$, Eq.\
(\ref{eq:diamond:m2:R-Oh})].  When the symmetry is reduced from $O_h
\times \gtis$ to $D_{4h} (D_{2d})$, the polar quasivector becomes
allowed, but the polar vector remains forbidden.  On the
other hand, the reduced symmetry $D_{2h} (C_{2v})$ implies that not
only a polar quasivector ($\Gamma_5^-$ of $O_h$) becomes allowed
[Eq.\ (\ref{eq:diamond:m2:D4h-C2v})], but also a polar vector
($\Gamma_4^-$ of $O_h$) becomes observable
\begin{equation}
  O_h \mapsto C_{2v} : \quad
  \Gamma_4^- \mapsto \Gamma_4 \oplus \Gamma_1 \oplus \Gamma_2 \,.
\end{equation}
It becomes clear from Table~\ref{tab:diamond:m2} that the reason for
the observability of both vectors under $D_{2h} (C_{2v})$ lies in
the fact that polar vectors and polar quasivectors are only
distinct quantities under high-symmetry point groups like $O_h
\times \gtis$ and $D_{4h} (D_{2d})$.  But they represent the same
observable physics when the symmetry is reduced to a group like
$D_{2h} (C_{2v})$ that makes both of these quantities measurable.
Under $D_{2h} (C_{2v})$, both quantities manifest themselves via
terms in the energy dispersion of band electrons proportional to the
invariant $k_x$ [Eq.\ (\ref{eq:diamond:m2:K-C2v})].

\subsection{Magnetization in diamond}
\label{sec:diamond:m1}

A magnetic dipole density representing ferromagnetic order in
diamond is analyzed in Table~\ref{tab:diamond:m1}.  Under the point
group $O_h$ of nonmagnetic diamond, the dipole density transforms
according to the IR $\Gamma_4^+$.  A magnetization pointing parallel
to the crystallographic direction $[001]$
[Fig.~\ref{fig:diamond-cub}(f)] reduces the symmetry to $D_{4h}
(C_{4h}) \rightarrow C_{4h}$.  As to be expected, the spin operator
$\sigma_z$ transforms according to $\Gamma_1^+$ of $C_{4h}$ and a
nonzero expectation value of $\sigma_z$ signals the presence of
ferromagnetic order.

\subsection{Multipolarization in Ga$_{1-x}$Mn$_x$As}
\label{sec:diamond:gammnas}

Ferromagnetic Ga$_{1-x}$Mn$_x$As and related (III,Mn)\=/V compounds
\cite{die14} are examples for multipolar materials.  Above the Curie
temperature, Ga$_{1-x}$Mn$_x$As has a zincblende structure
(Sec.~\ref{sec:diamond:e3}).  Below the critical temperature, it has
the magnetic space group $I\bar{4}m'2'$ (No.~119.319) and point
group $D_{2d} (S_4)$.  It inherits the electric octupolarization
($\ell = 3$) of the parent zincblende structure that manifests
itself via the Dresselhaus term (\ref{eq:diamond:e3:H}), i.e., this
material is electropolar.  Mn gives rise to a magnetopolarization
parameterized by a Zeeman-like exchange term
(Table~\ref{tab:diamond:m1}).  But Ga$_{1-x}$Mn$_x$As also supports
an antimagnetopolarization ($\ell = 2$) parameterized by a term as
in Eq.\ (\ref{eq:diamond:m2:K-D2d}) \cite{win20}.  In a TB model
\cite{cha75, cha77}, this term can be traced back to the fact that,
unlike the ferromagnetic structure in Fig.~\ref{fig:diamond-cub}(f), the
two sublattices of the diamond structure are distinct in
Ga$_{1-x}$Mn$_x$As.  Similarly, ferrimagnets are often multipolar.

\subsection{Correspondence between electric and magnetic order}
\label{sec:diamond:e-m}

Similar to lonsdaleite, electric and magnetic order in variants of
diamond follows a close correspondence.
Figures~\ref{fig:diamond-cub}(c) and~\ref{fig:diamond-cub}(d)
represent hexadecapolar and octupolar order due to local octupole
moments for both the electric and the magnetic case.  As summarized
in Table~\ref{tab:symmetries}, for odd $\ell$, the point group
characterizing the magnetopolar case is obtained from the group
characterizing the electropolar case by replacing space inversion
$i$ by time inversion $\theta$.

\section{Conclusions and outlook}
\label{sec:conclusions}

Using symmetry, we have developed a general theory of electric,
magnetic, and toroidal polarizations in crystalline solids.  We have
identified four families of multipole densities representing
even\=/$\ell$ and odd\=/$\ell$ electric and magnetic multipoles
(Table~\ref{tab:mpole-even-odd}).  Beyond the standard distinction
between electric and magnetic multipoles, each of these four
families bring about qualitatively different physics as they behave
differently under space inversion $i$ and time inversion $\theta$.
The four families of multipole densities give rise to five
qualitatively distinct categories of polarized matter
(Table~\ref{tab:mpole-even-odd}).  Even\=/$\ell$ electric multipole
densities may exist in all categories of polarized matter; they are
the only family of multipole densities permitted in parapolar media.
Electropolar, magnetopolar, and antimagnetopolar media permit each
exactly one other family of multipole densities, while multipolar
media permit all four families of multipole densities.  Each
category is characterized by distinct features in the band structure
of Bloch electrons (Fig.~\ref{fig:degen}).

Our group-theoretical analysis does not reference electromagnetism
to define multipolar order.  In this way, it avoids the difficulties
underlying an electromagnetic definition of multipole densities as
the multipole moment of an arbitrarily chosen unit cell.
Group theory is used, in particular, to derive the invariants
(\ref{eq:invars}) that incorporate the effect of electric and
magnetic multipolar order into the Hamiltonian for the dynamics of
Bloch electrons.  Nonetheless, in quantum-confined geometries, the
invariants (\ref{eq:invars}) reproduce the electromagnetic hallmarks
of electric and magnetic multipole densities including, e.g.,
equilibrium currents representing a magnetization ($\ell = 1$) or
magnetic quadrupolarization ($\ell = 2$), as demonstrated in a
recent study of the magnetoelectric effect in quasi-2D
systems~\cite{win20}.

Our analysis reveals that the familiar Rashba SO coupling
(\ref{eq:lons:e1:H}) represents the electric dipolarization in
electropolar materials such as wurtzite.  Rashba SO coupling has
been the starting point for countless fundamental studies and
applications that have greatly stimulated the field of
spintronics~\cite{man15, dja19}. Beyond that, our work establishes a
systematic correspondence between electric and magnetic multipolar
order and carrier dynamics (Table~\ref{tab:tensor-op-power} and
Fig.~\ref{fig:degen}) that provides a general framework for further
fundamental studies and a wide range of applications of multipolar
order in complex materials.

For example, antiferromagnetic order is often characterized as a
static spin density wave $\vek{S} (\vek{q})$ with a finite wave vector
$\vek{q}$ \cite{rad66, whi07}.  Similarly, antiferroelectric order
can be viewed as an electric dipolarization density wave $\vek{P}
(\vek{q})$ with $\vek{q} \ne 0$.  The electric and magnetic
multipole densities with $\ell \ge 1$ discussed here represent
macroscopic \emph{homogeneous} quantities corresponding to $\vek{q}
= 0$, like any field and material tensors characterized by Neumann's
principle \cite{nye85}.  Beyond spin density waves $\vek{S}
(\vek{q})$ and dipolarization density waves $\vek{P} (\vek{q})$
($\ell = 1$), we may envision electric and magnetic multipole
density waves $\vekc{m} (\vek{q})$ with $\ell > 1$ and $\vek{q} \ne
0$.  As it is the case for spin density waves $\vek{S} (\vek{q})$
\cite{rad66}, static waves $\vekc{m} (\vek{q})$ may, but need not, be
commensurate with the underlying crystal structure.  Electric and
magnetic order can be characterized by magnetic point groups if the
order is commensurate with the underlying crystal structure.
Polarization waves $\vekc{m} (\vek{q})$ for $\ell > 1$ can also
represent a generalization of magnons in ferromagnets \cite{whi07}
and ferrons in ferroelectrica \cite{bau22}.

\begin{table}
  \caption{\label{tab:space:subgroups} Maximal translationengleiche
  subgroups of the space group $D_{6h}^4$ of lonsdaleite and $O_h^7$
  of diamond \cite{hah05}, their crystal classes $G$, and the lowest
  orders $\ell_\mathrm{min}^{(\mathrm{e},g)}$
  ($\ell_\mathrm{min}^{(\mathrm{e},u)}$) of even (odd) electric
  multipole densities these space groups support \cite{kos63}.}
  \renewcommand{\arraystretch}{1.2} \let\mc\multicolumn
  \begin{tabular*}{\linewidth}{Ls{0.5em}ERLLCCl}
    \hline \hline
    \multicolumn{3}{c}{space group} & G & \rule{0pt}{3.0ex}
    \ell_\mathrm{min}^{(\mathrm{e},g)} &
    \ell_\mathrm{min}^{(\mathrm{e},u)} & \\ \hline \rule{0pt}{3.0ex}
    D_{6h}^4    &  194 & P6_3/mmc   & D_{6h} = D_6 \times \gsis & 2 &
    & lonsdaleite \\ \hline  \rule{0pt}{3.0ex}
    D_{3h}^4    &  190 & P\bar{6}2c & D_{3h} & 2 & 3 & \\
    D_{3h}^1    &  187 & P\bar{6}m2 & D_{3h} & 2 & 3 & \\
    C_{6v}^4    &  186 & P6_3mc     & C_{6v} & 2 & 1 & wurtzite \\
    D_6^6       &  182 & P6_322     & D_6    & 2 & 7 & (chiral) \\
    C_{6h}^2    &  176 & P6_3/m     & C_{6h} = C_6 \times \gsis & 2 & & \\
    D_{3d}^3    &  164 & P\bar{3}m1 & D_{3d} = D_3 \times \gsis & 2 & & \\
    D_{3d}^2    &  163 & P\bar{3}1c & D_{3d} = D_3 \times \gsis & 2 & & \\
    D_{2h}^{17} &   63 & Cmcm       & D_{2h} = D_2 \times \gsis & 2 & & \\
    \hline \rule{0pt}{3.0ex}
    O_h^7       & 227 & Fd\bar{3}m & O_h = O \times \gsis & 4 &
    & diamond \\ \hline  \rule{0pt}{3.0ex}
    T_d^2       & 216 & F\bar{4}3m & T_d    & 4 & 3 & zincblende \\
    O^4         & 210 & F4_132     & O      & 4 & 9 & (chiral) \\
    T_h^4       & 203 & Fd\bar{3}  & T_h    & 4 & 3 & \\
    D_{4h}^{19} & 141 & I4_1/amd   & D_{4h} = D_4 \times \gsis & 2 & & \\
    D_{3d}^5    & 166 & R\bar{3}m  & D_{3d} = D_3 \times \gsis & 2 & & \\
    \hline \hline
  \end{tabular*}
\end{table}  

We have illustrated our general theory by considering multipolar
order in crystal structures derived from lonsdaleite
(Sec.~\ref{sec:lons}) and diamond (Sec.~\ref{sec:diamond}).
Electric and magnetic multipole densities of different order $\ell$
yield crystallographic point groups as summarized in
Table~\ref{tab:symmetries}.  Generally, these multipole densities
reduce the crystal symmetry of pristine lonsdaleite and diamond.
To discuss the same physics from a different perspective, one can
start from the space group symmetry of the pristine material and
consider its different subgroups.  Table~\ref{tab:space:subgroups}
lists the maximal translationengleiche subgroups of the space group
of pristine lonsdaleite ($D^4_{6h}$, No.\ 194) and diamond ($O_h^7$,
No.\ 227) \cite{hah05}.  For each of these subgroups, we list the
associated crystal class $G$ and the lowest orders
$\ell_\mathrm{min}^{(\mathrm{e},g)}$
($\ell_\mathrm{min}^{(\mathrm{e},u)}$) of even (odd) electric
multipole densities \cite{kos63} permitted by these groups.  Beyond
the space groups already discussed in Secs.~\ref{sec:lons} and
\ref{sec:diamond}, this list includes also the space groups $D_6^6$
(No.\ 182, $P6_322$) and $O^4$ (No.\ 210, $F4_132$) that belong to
the chiral crystal classes $D_6$ and $O$, respectively.  Chiral
systems do not distinguish between polar vectors like wave vector
$\kk$ and axial vectors like spin $\vek{\sigma}$ \cite{bar04}.  Thus
it follows immediately that odd\=/$\ell$ electric multipole
densities in chiral systems manifest themselves via Dirac terms
$\propto \vek{\sigma} \cdot \kk$ (that may decompose into separate
terms $\propto \sigma_j k_j$ if the system is not cubic).

The group-theoretical tools underlying our analysis can be
integrated into crystallographic and materials databases to
facilitate materials research.  A systematic study of all 122
magnetic crystal classes will be published elsewhere.

\begin{acknowledgments}
  RW and UZ acknowledge stimulating discussions with R.~Resta and
  N.~Spaldin.  RW also benefited from discussions with A.~Hoffmann
  and M.~Norman.  Work at Argonne was supported by DOE BES under
  Contract No.\ DE-AC02-06CH11357.
\end{acknowledgments}


%%%%%%%%%%%%%%%%%%%%%%%%%%%%%%%%%%%%%%%%%%%%%%%%%%%%%%%%%%%%%%%%%%


\begin{thebibliography}{109}
\bibitem [{misc:cartesian-tensors()}]{misc:cartesian-tensors}%
  \BibitemOpen
  \bibinfo {note} {Electric and magnetic multipoles can also be viewed as
  cartesian rank-$\ell$ tensors \cite{jac99, raa05}. However, as is well-known
  \cite{edm60}, spherical rank-$\ell$ tensors transform irreducibly under the
  rotation group $\Rit$, whereas cartesian tensors of rank $\ell \ge 2$ are
  reducible, which is the reason why we prefer to define electric and magnetic
  multipoles as spherical tensors and not as cartesian tensors.}\BibitemShut
  {Stop}%
\bibitem [{misc:incommensurate()}]{misc:incommensurate}%
  \BibitemOpen
  \bibinfo {note} {This classification is not exhaustive, see
  Sec.~\ref{sec:conclusions}.}\BibitemShut {Stop}%
\bibitem [{misc:iv-space-groups()}]{misc:iv-space-groups}%
  \BibitemOpen
  \bibinfo {note} {Systems transforming according to so-called type-IV magnetic
  space groups \cite{bra72} possess a black-and-white Bravais lattice. These
  are antiferromagnetic systems that belong to a nonmagnetic crystal class
  \cite{cor58a, bir64}, and they are parapolar or electropolar according to the
  present classification. All systems studied in the present work have type-II
  or type-III magnetic space groups.}\BibitemShut {Stop}%
\bibitem [{misc:double()}]{misc:double}%
  \BibitemOpen
  \bibinfo {note} {The isomorphisms between magnetic and nonmagnetic
  crystallographic point groups exist at the level of the single (spinless)
  groups relevant, e.g., for Neumann's principle. Only at the level of double
  groups, these groups are generally not isomorphic because then $i^2 = e$,
  whereas $\theta^2 = \bar{e}$ with $\bar{e}^2 = e$.}\BibitemShut {Stop}%
\bibitem [{misc:compat-nonunique()}]{misc:compat-nonunique}%
  \BibitemOpen
  \bibinfo {note} {The compatibility relations relating two crystallographic
  point groups $G$ and $U$ are not uniquely defined. They depend on the
  relative orientation of the coordinate systems used to define the symmetry
  operations in $G$ and~$U$.}\BibitemShut {Stop}%
\bibitem [{misc:struct-phase-trans()}]{misc:struct-phase-trans}%
  \BibitemOpen
  \bibinfo {note} {These considerations can be extended to include, e.g., the
  possibility that a magnetic phase transition may be accompanied by a
  structural phase transition. If a nonmagnetic structure undergoes a phase
  transition to a magnetic phase, by definition a magnetic multipole becomes
  allowed. The phase transition may likewise affect the electric order. For
  example, it is well-known that iron is cubic above the Curie temperature and
  therefore has an electric hexadecapolarization ($\ell = 4$), but it develops
  a tetragonal distortion in the ferromagnetic state \cite{whi07} that allows
  an electric quadrupolarization ($\ell = 2$).}\BibitemShut {Stop}%
\bibitem [{misc:axions()}]{misc:axions}%
  \BibitemOpen
  \bibinfo {note} {This discussion also applies to magnetic monopoles and
  axions transforming according to $D_0^{--}$ of $\Rit$. These quantities are
  forbidden by $\Rit$ in vacuum but are symmetry-allowed by 32 of the 122
  magnetic crystallographic point groups.}\BibitemShut {Stop}%
\bibitem [{misc:lons-indicator()}]{misc:lons-indicator}%
  \BibitemOpen
  \bibinfo {note} {Beyond the invariant (\ref{eq:lons:e2:H}), the Hamiltonian
  for Bloch electrons in lonsdaleite also includes a spherically symmetric
  invariant that transforms according to $D_0^+$ of $R_i$ so that it transforms
  according to $\Gamma_1$ of any subgroup of $R_i$ \begin{displaymath}
  H^\inv{e}{0} = a^\inv{e}{0} \mathcal{m}^\inv{e}{0} (k_x^2 + k_y^2 + k_z^2)
  \,. \end{displaymath} Thus, instead of $H^\inv{e}{0}$ and $H^\inv{e}{2}$, we
  could also choose for $D_{6h}$ the linearly independent invariants
  \begin{displaymath} H^\inv{e}{2}_{xy} = b_{xy}^\inv{e}{2} (k_x^2 + k_y^2)
  \,\, , \qquad H^\inv{e}{2}_{z} = b_{z}^\inv{e}{2} \, k_z^2 \,,
  \end{displaymath} with \begin{align*} b_{xy}^\inv{e}{2} & = a^\inv{e}{0}
  \mathcal{m}^\inv{e}{0} + a^\inv{e}{2} \mathcal{m}^\inv{e}{2} \\
  b_{z}^\inv{e}{2} & = a^\inv{e}{0} \mathcal{m}^\inv{e}{0} - 2 a^\inv{e}{2}
  \mathcal{m}^\inv{e}{2} \,. \end{align*} However, both $k_x^2 + k_y^2$ and
  $k_z^2$ already have nonvanishing expectation values under $R_i$, whereas the
  expectation value of $K^\inv{e}{2}_1$ vanishes under $R_i$. Therefore
  $I^\inv{e}{2}_1 \propto K^\inv{e}{2}_1$ constitutes the indicator for the
  electric quadrupole density in lonsdaleite.}\BibitemShut {Stop}%
\bibitem [{misc:diamond-afm()}]{misc:diamond-afm}%
  \BibitemOpen
  \bibinfo {note} {The configuration shown in Fig.~\ref{fig:diamond-cub}(e) is
  the classic diamond-lattice antiferromagnet \cite{oit18}. Another example of
  magnetic quadrupolar order is discussed in Ref.\ \cite{hay18a}.}\BibitemShut
  {Stop}%
\end{thebibliography}
\end{document}